\documentclass[]{aastex63}

\received{}
\revised{}
\accepted{}
\submitjournal{ApJ}

\shorttitle{Carbon in compact Galactic PNe}
\shortauthors{Stanghellini, et al.}
\graphicspath{{./}{figures/}}

\begin{document}

\title{Carbon Abundances in Compact Galactic Planetary Nebulae: An Ultraviolet spectroscopic study with the {\it Space Telescope Imaging Spectrograph (STIS)}.}

\correspondingauthor{Letizia Stanghellini}
\email{letizia.stanghellini@noirlab.edu}

\author[0000-0003-4047-0309]{Letizia Stanghellini}
\affiliation{NSF's NOIRLab,
950 N. Cherry Ave., Tucson, AZ 85719, USA}

\author[0000-0001-5865-4953]{Rafia Bushra}
\affiliation{School of Industrial and Systems Engineering,
University of Oklahoma,
202 W. Boyd St., Room 124,
Norman, OK 73019, USA}

\author[0000-0003-4058-5202]{Richard A. Shaw}
\affiliation{Space Telescope Science Institute,
3700 San Martin Drive,
Baltimore, MD 21218, USA}

\author[0000-0003-2442-6981]{Flavia Dell'Agli}
\affiliation{INAF-Osservatorio Astronomico di Roma, Via Frascati 33, 00078, Monte Porzio Catone (RM), Italy}
\author[0000-0002-1693-2721]{D. A. Garc{\'\i}a-Hern{\'a}ndez}
\affiliation{Instituto de Astrof{\'\i}sica de Canarias (IAC), E-38205 La Laguna, Tenerife, Spain}
\affiliation{Universidad de La Laguna (ULL), Departamento de Astrof{\'\i}sica, E-38206 La Laguna, Tenerife, Spain}
\author[0000-0002-5026-6400]{Paolo Ventura}
\affiliation{INAF-Osservatorio Astronomico di Roma, Via Frascati 33, 00078, Monte Porzio Catone (RM), Italy}

\begin{abstract}

We surveyed a sample of compact Galactic planetary nebulae (PNe) with the {\it Space Telescope Imaging Spectrograph} on the {\it Hubble Space Telescope} ({\it HST}/STIS) to determine their gas-phase carbon abundances. Carbon abundances in PNe constrain the 
nature of their asymptotic giant branch (AGB) progenitors, as well as cosmic recycling.
We measured carbon abundances, or limits thereof, of 11 compact Galactic PNe, notably increasing the sample of Galactic PNe whose carbon abundance based on {\it HST} ultraviolet spectra is available.
Dust content of most targets has been studied elsewhere from Spitzer spectroscopy; given the compact nature of the nebulae, both UV and IR spectra can be directly compared to study gas- and dust-phase carbon. We found that carbon-poor (C/O$<$1)  compact Galactic PNe have oxygen-rich dust type (ORD), while their carbon-enhanced counterparts (C/O$>$1) have carbon-rich dust (CRD), confirming the correlation between gas- and dust-phase carbon content which was known for Magellanic Cloud PNe. 
Based on models of expected final yields from AGB evolution we interpret the majority of the carbon-poor PNe in this study as the progeny of $\sim 1.1-1.2$~M$_{\odot}$ stars that experienced some extra-mixing on the red giant branch (RGB), they went through the AGB but did not go through the carbon star phase. Most PNe in this group have bipolar morphology, possibly due to the presence of a sub-solar companion. 
Carbon-enhanced PNe in our sample could be the progeny of stars in the $\sim 1.5-2.5$ M$_{\odot}$ range, depending on their original metallicity.

\end{abstract}
\keywords{ISM: abundances, abundances, planetary nebulae: general}

\section{Introduction} \label{sec:intro}
Planetary nebulae (PNe), the gas and dust remnants of low- and intermediate-mass stars (LIMS, $\sim1<{\rm M/M_{\odot}}<$8), are key probes of the chemical evolution in galaxies. Understanding how LIMS evolve is extremely important in astrophysics since the LIMS population represents most of the stellar mass in galaxies. Most LIMS are believed to go through the asymptotic giant branch (AGB) phase, which prominently contributes to the integrated luminosity of a galaxy \citep[e.g.,][]{2011ASPC..445..391M}. Furthermore, AGB stars and the subsequent PNe are major dust producers. It is thus important to have 
the best observational data sets to constrain the nucleosynthesis models at various metallicities.

At the end of their lives, LIMS become major producers of C and N. Nucleosynthesis of these elements occurs in the stellar core. Mass-loss brings these elements to the ISM after they are dredged up to the stellar surface. Planetary nebula abundances of the major elements are typically straightforward to measure 
by analyzing emission lines. While ground-based telescopes allow the direct observation of most key elements in Local Group PNe, carbon remains elusive since its major collisionally excited and bright emission lines,
\ion{C}{2}] $\lambda$2325-29, \ion{C}{3}] $\lambda$1909, and \ion{C}{4} $\lambda$1550 \AA, are emitted in the satellite ultraviolet. Carbon recombination lines are emitted in the optical spectrum, and they are much fainter than the collisionally excited lines in the UV; combined with nearby oxygen emission features, they are useful to constrain the C/O ratio \citep{GR18}.
Yet carbon is an essential element: carbon and its compounds relate to the origin of life in the Universe, which makes it fundamental to understand where it forms and how 
its abundance grows over time. Furthermore, stellar evolution theory 
predicts that processes defining final yields of nitrogen and carbon, such as third dredge-up (TDU) and hot bottom burning (HBB), strongly depend on the progenitor mass, thus carbon concentration in PNe (especially relative to nitrogen) is a signature of the mass-range of the progenitor LIMS, thus, ultimately, of their age.

To understand dust formation and evolution in the context of stellar and galactic evolution, \citet{SGG07} have observed dust features in Magellanic Cloud PNe from the {\it Spitzer Space Telescope Infrared Spectrograph} (Spitzer/IRS) spectra, and gas-phase PN properties from {\it Hubble Space Telescope (HST)} imaging and UV spectroscopy, and found that nebular gas chemistry, dust  composition, and PN morphology are correlated. The IRS spectra carry a wealth of information on the dust continuum and solid-state dust features. Intermediate-mass LIMS in the Magellanic Clouds produce symmetric (i.e., non-bipolar), carbon-rich PNe with carbon-rich dust features (such as polycyclic aromatic hydrocarbons (PAHs), hydrogenated amorphous carbon grains (HACs), etc.), while high-mass progenitors produce generally bipolar, nitrogen-rich, PNe with oxygen-rich dust (e.g., amorphous and crystalline silicates). 

We have learned from the Magellanic Cloud PN project that the simultaneous availability of IRS spectra, narrow and broad-band {\it HST} images, optical, and UV spectra, yields detailed insight into the post-AGB and PN evolution. We need to build a similar data set for compact Galactic PNe -- PNe whose maximum angular radii are smaller than $\sim$4$-$5 \arcsec-- to extend the analysis across metallicities. 
Compact  Galactic PNe, defined and analyzed by \citet{SSV16}, have many advantages with respect to extended PNe when used as evolutionary probes. The relevant property that makes compact PNe compelling in this study is that their spectra, from UV to optical to IR, can be acquired with just one pointing to include the whole nebula, which in turn provides plasma diagnostics and chemical analysis of the PN as a whole. 
Their compact shapes allowed the study of dust content with Spitzer/IRS spectroscopy and with other spectroscopy without the problem of aperture correction. Furthermore, compact Galactic PN spectra, no matter whether UV, optical, or IR, are analyzed identically in the Galactic samples and in the Magellanic Clouds, thus the comparative analysis of the various samples is direct and unambiguous.

A detailed search through the literature disclosed that only a few Galactic PNe with Spitzer/IRS spectroscopy have reliable gas-phase carbon abundances \citep{VSD17}. We thus embarked on this spectroscopic study of carbon in compact Galactic PNe. Our observing goals were to acquire a sizable set of UV spectra to detect strong UV carbon transitions of PNe whose Spitzer/IRS spectra were also available, and whose nitrogen, oxygen, and other elemental abundances were available in the literature, to measure their gas-phase carbon content, and to study their correlation with the dust-phase carbon and, in the context of PN progenitors and their evolution, to compare their surface chemistry with the final surface chemical abundances of AGB stars, with the final goal of constraining mass and metallicity (and age) of the progenitors. The {\it HST} is the only telescope, and the {\it Space Telescope Imaging Spectrograph (STIS)} \citep[STIS:][]{STIS, STIS_OOP} 
the best instrument, that can be used to measure carbon abundances in compact Galactic PNe.

This paper represents the first systematic study of carbon abundances -- from direct observations of UV lines -- of compact Galactic PNe with known dust-phase chemistry. With this study, we considerably augmented the observational data with which to constraint AGB evolution in the Galaxy.
Prior to this study, there were only 7 Galactic PNe, compact or otherwise, whose {\it HST} UV spectra could be employed to determine the gas-phase abundance of carbon \citep{Henry2015, 2015ApJ...803...23D} in addition to \citet{Henry2008} observations of the halo PN DdDm 1 (PN~G061.9+41.3). Another ~$\sim$30 Galactic PNe had been previously observed with the {\it International Ultraviolet Explorer} (IUE), providing reliable carbon abundances mostly for nearby, extended PNe \citep[and references therein]{VSD17}. Dust and gas-phase carbon properties have been studied by \citet{DIR14}, based on a sample of mostly extended Galactic PNe, and with aperture corrections applied to several targets.
Finally, as a comparison, carbon abundances are available for 11 Small Magellanic Cloud (SMC) and 24 large Magellanic Cloud (LMC) PNe, all from UV emission lines and observed with the {\it HST} \citep{Stanghellini05, Stanghellini09}.

\section{Observing Program} \label{sec:Program}
\subsection{Observations} \label{sec:Observations}
Our observations were obtained in {\it HST} program GO--15211, which was extended by the mission in program GO--16013, with observations taking place between 2018 Jan 05 and 2020 Sept 20. 
We selected our targets to be spatially compact Galactic PNe (with apparent radii $\theta\leq$5\arcsec), preferentially already observed in the optical wavelengths with {\it HST} (e.g., program GO-11657), which, in addition to providing their size and morphology, greatly simplifies target acquisition. 
We observed each target with FUV/G140L and NUV/G230L spectroscopic configurations with STIS. 
The aperture was placed on the center of each nebula to detect the central stars (CSs) as well. 
The program is not dissimilar from the UV {\it HST} program targeting LMC PNe (GO-9120), since PNe in the LMC have similar maximum extensions to our compact Galactic PNe. 
Our allocation of 75 targets in Cycles 25 and 26 was only partially fulfilled, as expected in ``snapshot'' mode. 
Because we were primarily interested in obtaining total fluxes in critical lines of C, in order to facilitate a direct comparison with ground-based optical observations, each target was observed with the $6\arcsec\times6\arcsec$ aperture. 
As all or most of the flux from these targets is emitted within about 5\arcsec\ \citep{SSV16}, this aperture is nearly equivalent to slitless spectroscopy and carries the advantage of excluding bright UV sources in the field which could pose a risk to the MAMA detectors. 
This choice comes at the cost of greatly diminished spectral resolution, which is set primarily by the angular extent of each target. 

Our observing plan specified exposures with both the FUV-MAMA detector with the low-resolution grating G140L and the NUV-MAMA with G230L.
The observing log is presented in Table~\ref{tab:ObsLog}. 
Observations for three targets failed due to instrument or telescope problems,
but they are included in Table~\ref{tab:ObsLog} (with zero observing time) for completeness. Also notable, PN~G286.0--06.5 was observed in both programs.

We planned exposure times to achieve a signal-to-noise (S/N) ratio $>10$ in the brightest emission lines of carbon over the extent of the target. This is sufficient to obtain good C elemental abundances since one or more of the observable ions C$^+$, C$^{+2}$, and C$^{+3}$ will dominate the emission. 
The exposure durations were however limited to a maximum of 1200~s to ensure that both FUV and NUV spectra could be obtained for each target within a single orbit; single-orbit visits are required for ``snapshot" programs. 

All data analyzed for this program are available in the Mikulski Archive for Space Telescopes (MAST) at the Space Telescope Science Institute. The specific observations we analyzed can be accessed via \dataset[https://doi.org/10.17909/t9-n5ef-9894]{https://doi.org/10.17909/t9-n5ef-9894}.

\subsection{Data Reduction and Spectrum Extraction} \label{subsec:reduction}

We reduced the data through flat-field correction with the distributed python package \textbf{stistools}, which embodies the STScI CALSTIS calibration pipeline as a library. 
Our observations used an available-but-unsupported observing mode, which obligated us to rectify the flattened images and extract the one-dimensional spectra with custom software. 
The raw data were first processed with the CALSTIS task \textit{basic2d} to perform 2d image reduction. 
To initialize the data quality array, \textit{basic2d} uses a bad pixel reference table and performs a bitwise OR with the initial data quality file. This routine appropriately combines data quality information from neighboring pixels before performing the OR operation in order to take Doppler smearing and binning into account. 
The primary cause for dark current in MAMA detectors is believed to be a phosphorescent glow from impurities in the detector window. In short time scales, the glow varies exponentially with temperature; in a long time scale, the behavior becomes more complex. With \textit{basic2d} we subtracted the dark signal using relevant reference files and updated the science data quality files for bad pixels in the dark reference file. 
Lastly, we used \textit{basic2d} to correct for pixel-to-pixel and large-scale sensitivity gradients using a p-flat and l-flat reference file respectively. P-flats are configuration-(grating, central wavelength, detector, etc.) dependent flat field images, with no large-scale sensitivity variation. L-flats are sub-sampled flat field images that contain large-scale sensitivity variation across the detector. \textit{basic2d} combines these two types of flat field and then corrects the science image by dividing it with the combined flat field image. The data quality and error arrays are updated again to account for flat-fielding.

To perform spectral extraction, we need to consider that
the spectral trace orientation on STIS detectors slowly changes over time and is also subject to an offset that is unique to each observation caused by the MSM (mode selection mechanism) positioning. To rectify this we use the IRAF task \textit{mktrace}, which corrects the orientation of spectral trace, re-centers it on a new row, and provides the new center as an output. However, it takes an approximate center as an input. 
To determine our best approximate center, we fit a 1d Gaussian model to a column with a strong signal from the trace. We also use this Gaussian model to determine approximate box parameters, used in our last extraction step. Although the Gaussian fitting does provide approximate box parameters, we did adjust these manually on the actual image. 

The \textit{x2d} task then rectifies the image using bi-linear point interpolation. For each point in the rectified output image, there is a corresponding point in the distorted input image and the four nearest pixels are bi-linearly interpolated to determine the value to be assigned to the point in the output image. 
This mapping from output pixel to input images is done by using the dispersion relation 
and the spectral trace table generated by \textit{mktrace}. 
Pixel number as a function of wavelength is given by the dispersion relation and the displacement in cross-dispersion direction at each pixel along the dispersion direction is given by the trace table. Appropriate corrections are applied to take binning into account.
Lastly, the \textit{x2d} task converts the counts to surface brightness in [ergs cm$^{-2}$ sec$^{-1}$ \AA$^{-1}$ arcsec$^{-2}]$. 

Examples of the 2d spectra are shown in the upper and middle panels of Figs.~1 through~4. The 2d spectra clearly show the footprint of the nebular shape as observed in correspondence of the major nebular emission lines. For example, the elliptical (Figs. 1, 2, and 3) vs. bipolar (Fig. 4) shapes are clearly identified for most individual spectral images. The 2d spectra also show the presence of stellar continua. 

After masking the bad pixel in the 2d image, we extract the 1d spectra using our own python routine. We chose a spectral extraction box as large as the largest feature in the 2d spectral image, and we subtract the average background from two regions on either side of the spectral trace. If the PN had been previously observed with the {\it HST} cameras -- i.e., it has been spatially resolved -- we use the measured photometric diameter as a guide to the initial guess for the extraction box size.

The final spectra have been calibrated through the default wavelength calibration since we did not observe simultaneous comparison arcs. This is not an issue for UV spectra of Galactic PNe, where the major emission lines are easily recognizable, as we can see in the 2d spectral images. We thus adjusted the zero-point of the wavelength solution of each extracted 1d spectra based upon the brightest known nebular emission features. 
Figs. 1 through 4 (lower panels) also show the extracted 1d spectra of the PNe. 

 \section{Analysis} \label{sec:analysis}

\subsection{Emission Line Measurements} \label{subsec:fluxes}

The rectified 2d spectrograms and the extracted 1d spectra of the observed PNe have been inspected for features. For the following, we discuss only targets that show at least one nebular emission line, with sufficient S/N for the subsequent analysis. We list these PNe in Table~\ref{tab:Parameters}. Table 2 also gives the ancillary parameters that are relevant for our analysis and include the distance from the Galactic plane, the He, N, and O atomic abundances, the nebular morphology, and the dust type derived from IRS spectral analysis. All parameters are referenced in the Table.

We measured the emission-line fluxes with the IRAF task {\it splot}. The emission line intensities have been measured via Gaussian fit, which is a good model for all the measured lines. The flux uncertainties given for the emission lines are the random error estimates assuming the Gaussian shape. We checked that this approximation worked for all of the lines used for abundance measurements. Naturally, this procedure assumes, as it is often the case in spectroscopic analysis, that the continuum level is well identified. If the continuum was mismatched, there would be additional uncertainty, up to about 10$\%$, for the flux lines.
All fluxes and their uncertainties, derived from the above measurement procedure, were scaled to $I_{\rm H\beta}=100$. 
The observed intensities and extinction-corrected fluxes are related by:
\begin{equation}
    \frac{I_{\rm \lambda}}{I_{\rm \beta}} = \frac{F_{\rm \lambda}}{F_{\rm \beta}} 10^{cf_{\rm \lambda}} \times 100
\end{equation}
Here, $c$ is the logarithmic extinction at $H_{\beta}$ for each target, and $f_\lambda$ is the wavelength dependent reddening function from \citet{Cardelli1989}. 
We obtain the final line intensities by using the H$_{\beta}$ fluxes and extinction constants from \citet{SSV16}, except for PN~G001.6+8.3 and PN~G281.0-05.6, whose parameters are from \citet{CKS}.
The measured line fluxes and intensities, and the $H_{\beta}$ fluxes and extinction constants, are given in Tables~\ref{tab:Flux1}--\ref{tab:Flux4}. 
More details on individual nebular spectra are given in Section \ref{individual}. 

\subsection{Plasma Diagnostics} \label{subsec:diag}

The electron densities ($N_{\rm e}$) and temperatures ($T_{\rm e}$) adopted for the abundance calculation are given in Table~\ref{tab:Diag}. Most of the diagnostics have been taken from the literature (cited within the Table), since the UV ranges observed in this work do not include any diagnostic lines. 
When both T$_{\rm e}$[\ion{N}{2}] and T$_{\rm e}[$\ion{O}{3}] where available for the same PN, we used T$_{\rm e}$[\ion{N}{2}] for the C${^+}$ abundances, and T$_{\rm e}[$\ion{O}{3}] for the C$^{2+}$ and C$^{3+}$ abundance calculation. We always used the [\ion{S}{2}] densities preferentially, if available for a given PN. When the electron density or temperature was not available, we calculated them from published diagnostic line intensities \citep{AMO92} using the \textit{pyneb} package in python \citep{LMS15}.
In one case we could not find plasma diagnostics nor diagnostic flux ratios in the literature, thus we adopted a typical values of $N_{\rm e}$ to calculate abundances, as noted in Table~\ref{tab:Diag}. 

\subsection{Abundance Analysis}
\label{subsec:abundanalysis}

We measured ionic abundances from the line fluxes and ancillary diagnostics with the \textit{pyneb} package, and the atomic data set therein, as also given in Table 8.  All ionic abundances are intended in terms of H$^{+}$; heretofore, we use the term ''ionic abundance" to mean ''ionic abundance ratios to H$^{+}$". 
The derived abundances of the carbon ions are presented in Table 9; ionic abundances of other elements derived from the same UV spectra are in Table 10.  

Uncertainties in the line intensity and plasma diagnostics both contribute to the final abundance errors. 

We measured the uncertainties in the ionic abundances due to line uncertainties (including reddening correction) by Monte-Carlo simulation assuming Gaussian distribution centered on the measured intensities. The resulting uncertainties are in the 0.001--0.04 dex range for the ionic abundances.

The contribution to the abundance uncertainties from the line ratio are the only ones that we can actually calculate with our data set. On the other hand, by far the dominant source of uncertainty in PN abundances stems from plasma diagnostics. We adopted electron density and temperature values from the literature, and none of the original references in Table~\ref{tab:Diag} give uncertainties for these parameters. 
We thus estimate the magnitude of final abundance uncertainty as it stems from guessed 5$\%$ and 10$\%$ uncertainties in both the electron density and temperature. Temperature shifts due to inhomogeneity in the atomic data sets, e.g. \citet{JDD}, are folded in these assumed values.

We found that the electron density uncertainty has no effect of the final ionic abundances. In fact, for all ions and all PNe, a 10$\%$ uncertainty in the density produces ionic abundance uncertainties in the 1$\times$10$^{-3}$--3.5$\times$10$^{-5}$ dex range. On the other hand, a 10$\%$ uncertainty in the T$_{\rm e}$ propagates into a mean uncertainty of 0.39$\pm$0.03 dex in the ionic abundances. The assumption of $\Delta T_{\rm e}\sim 10\%$ is conservative, while a $\pm$5$\%$ uncertainty is more realistic, and translates into a 0.17$\pm$0.01 dex uncertainty in the final abundances. 
The final uncertainties are thus dictated by our guesses of the electron temperature uncertainties. Since these initial guesses translate into abundance uncertainties in such narrow ranges, we will use their averages as final uncertainties. 

It is worth noting that a formal analysis would yield to a final C/H abundances uncertainty which is $\sqrt(N)$ times the ionic uncertainty adopted, where N is the number of available ions.

We calculated atomic carbon abundances from the ionic values, using the scheme by \citet{KB94} to correct for unobserved ionization stages. In our spectra we expect to see the transitions relative to \ion{C}{2}], \ion{C}{3}], and \ion{C}{4}. 
In the section below, where applicable, we describe how the atomic carbon abundance has been derived, and which literature data we used to derive the Ionization Correction Factor for carbon (ICF(C)). Both ionic and atomic carbon abundances, and the ICF's, are listed in Table~\ref{tab:Carbon}.

\section{Individual PNe}\label{individual}
\subsection{PN~G003.9--14.9}
The 2d STIS UV spectrograms spatially resolve this PN for the first time (this PN was previously unobserved with {\it HST}). We found that an extraction box of 2\arcsec\ encompasses most of the spectra in both gratings. The nebular has an elliptical shape, although the morphology determination is uncertain on the UV spectra. The G140L image shows a very good spectral trace with high S/N; it shows the \ion{C}{4} emission feature with a P-Cygni profile, whose spatial extension indicates its stellar origin. We could not detect \ion{He}{2} $\lambda$1640 \AA~ in the G140L spectrum, even if a faint emission feature corresponding to \ion{He}{2} $\lambda$4686 has been detected in the optical spectrum \citep{1994A&AS..106..559T}. The P-Cygni feature with emission centered around $\lambda$1232 \AA~ could be \ion{N}{5} $\lambda$1239--43 \AA, similar to what was observed in the LMC PNe SMP~18 and SMP~25 (Stanghellini et al. 2005). The G230L spectrum does not show emission lines. 

\subsection{PN~G006.1+08.3}
A box of 2\arcsec\ encompasses the nebular UV spectrum of this Galactic Bulge, roughly elliptical PN, previously unobserved with {\it HST}. This PN had not been previously spatially resolved from the ground \citep{2003A&A...405..627T}. Our STIS program acquired only the G230L spectrum, showing a (likely) \ion{C}{3}] emission line at $\lambda$1890~\AA~($\Delta\lambda\sim-17$). Another notable feature at $\lambda\sim$2810~\AA~ is unidentified. We note a strong nebular continuum for $\lambda>2300$ \AA. The stellar spectrum is prominent, although an estimate of the CS temperature is problematic, given the lack of the G140L spectrum. We could not find the optical line's strengths to estimate the ICF to correct for the unobserved \ion{C}{2}] line, thus the atomic carbon abundance in Table~\ref{tab:Carbon} is a lower limit thereof. 

\subsection{PN~G025.3--04.6}
The STIS G230L spectrum shows nebular continuum emission and a likely emission feature which could be identified as \ion{C}{3}] at $\lambda$1907--09 \AA. The G140L spectrum does not show obvious emission lines. The atomic carbon abundance in Table~\ref{tab:Carbon} has been derived from 
C=ICF(C)$\times$C$^{2+}$.  We used the fluxes by \citet{GHG14} to derive the oxygen abundances needed to estimate the ICF.
 
 \subsection{PN~G038.4--3.3}
 Both STIS spectra of this PN are noisy. We were unable to make unambiguous line identifications.
 
 \subsection{PN~G042.9--06.9}
  The only emission feature observed for this PN is in the G230L spectrum, which we interpret to be \ion{C}{3}] at $\lambda$1907-09~\AA.  We used the optical emission lines from \citet{GHG14} to correct for the unseen emission lines via ICF analysis, as in PN~G025.3--04.6. 
 
 \subsection{PN~G053.3+24.0}
 The \ion{He}{2} feature at $\lambda$1640 \AA~in the G140L spectrum is very strong, which is indicative of a medium to high excitation PN. The only other feature in the G140L spectrum is a very faint, noisy emission at $\lambda\sim$1243~\AA~that could be \ion{N}{5}. The flux of the emission line identified as \ion{C}{3}] at $\lambda$1907-09 \AA~in the G230L spectrum corresponds to a very broad feature, which can be used as an upper limit to the emission line flux. 
We also list in Table 4 a couple of possible \ion{O}{3} features, although the wavelengths of the identified lines are not perfect matches to those observed. This is not a low-excitation PN, thus we did not apply the ICF corrections to measure the atomic carbon abundance, following \citet{KB94}'s prescription.
  
 \subsection{PN~G068.7+14.8}
The \ion{C}{4} emission has a P-Cygni profile and is not spatially extended, while the \ion{C}{2}] and \ion{C}{3}] carbon lines show spatial extension. We infer that the former is of stellar origin.  Since the optical $\lambda$4686\AA~ emission line has been detected \citep{1994A&AS..106..559T}, this could be a medium to high excitation PN. We thus derive the abundance without ICF, as in PN~G053.3+24.0. The G140L spectrum shows a faint line emission at $\lambda\sim$1242~\AA, which could be \ion{N}{5}, possibly of stellar origin as well. 

\subsection{PN~G097.6--02.4}
Both UV STIS spectra are very noisy, and we were unable to make unambiguous line identifications.
 
\subsection{PN~G107.4--02.6}
The STIS G140L spectrum is too noisy for line detection, and the \ion{C}{3}] emission line is the only line detected in the G230L spectrum. There is no information in the literature about the low-excitation \ion{O}{2} lines thus we can not calculate the ICF to get the total carbon abundance. As a result, the carbon abundance in Table~\ref{tab:Carbon} is a lower limit thereof. This is a similar case to PN~G006.1+08.3.

\subsection{PN~G232.8--04.7}
Both UV STIS spectra are too noisy for line identification.

\subsection{PN~G264.4--12.7}
Both the \ion{C}{4} and the \ion{N}{5} features in the G140L spectrum have P-Cygni profiles, and they both look stellar in origin. The \ion{C}{2}] and \ion{C}{3}] lines are very faint and extended in the G230L spectrum, and their abundances could not be measured.

\subsection{PN~G275.3--4.7}
The presence of the \ion{He}{2} in the G140L spectrum indicates a medium to high excitation PN. In order to correct for the missing \ion{C}{2}] line we use the optical oxygen lines from the literature \citep{2002ApJS..138..285M}, from which we estimate ICF(C)=1.037. 

\subsection{PN~G278.6--06.7}
The G140L spectrum is characterized by strong \ion{He}{2} and \ion{C}{4} emission lines from the whole volume of the PN, thus indicating a medium to high excitation PN, which agrees with the presence of the optical emission line at $\lambda$4686~\AA, corresponding to \ion{He}{2} emission \citep{1994A&AS..106..559T}. There is a noisy emission line corresponding to \ion{N}{5} in the G140L spectrum as well. 
 As clearly seen in Figure~3, the spatial distribution of the \ion{C}{3} and the \ion{C}{2} emission shows ionization stratification.
The atomic carbon abundance is the sum of the measured ionic abundances.

\subsection{PN~G281.0--05.6}
The \ion{C}{3}] line is the only emission line detected in the G230L spectrum and presents itself as a very broad emission. We correct for the undetected \ion{C}{2}] emission from the optical oxygen lines in the literature \citep{2002ApJS..138..285M}, finding ICF(C)=1.084, which has been used to calculate the total atomic abundance in Table~\ref{tab:Carbon}. There are P-Cygni lines corresponding to \ion{N}{5} and \ion{C}{4} in the STIS G140L spectrum. Their extensions indicate that they are probably of stellar origin. 

\subsection{PN~G286.0--06.5}
  There is a very noisy emission, not measured, that could be \ion{He}{2} at $\lambda$1640~\AA. There are \ion{N}{5} and \ion{C}{4} emission lines in the G140L spectrum with P-Cygni profiles; their extension indicates that they are probably of stellar origin. The atomic carbon abundance has been calculated by the sum of the \ion{C}{2} and \ion{C}{3} abundances.
 
\subsection{PN~G295.3--09.3}
In the G140L spectrum, the emission line corresponding to \ion{N}{5} has P-Cygni profile, and the emission flux given has high uncertainty given the shape of the underlying continuum. The identification of the [\ion{O}{3}] line is uncertain. The only nebular carbon line detected here is \ion{C}{3}]. There are no lower excitation transition intensities available in the literature for this PN to correct for \ion{C}{2}], thus the atomic abundance in the Table is a lower limit thereof.

\subsection{PN~G351.3+07.6}
 Both \ion{N}{5} and \ion{C}{4} emissions in the G140L spectrum have P-Cygni profiles. There are no emission features detected in the G230L spectrum. This is a similar case to PN~G003.9--14.9 and PN~G264.4--12.7.

\section{Comparison between PN abundances and stellar evolutionary models}

In this section, we characterize the observed PN sample in the framework of stellar evolution of LIMSs. The chemical abundances of PNe reveal the nucleosynthesis and mixing processes experienced by the star during the previous evolutionary phases. 
Before entering the PNe stage, stars of masses 0.8$\leq$M/M$_{\odot}\leq$8 experience H- and He- shell burning phases while climbing along the AGB \citep{sh65,Iben1975,Iben1976}. 

The low-mass threshold ($0.8~M_\odot$) is partly dependent
on the description of mass-loss adopted, as a more efficient mass-loss during the red giant branch (RGB) and the phases following the
core helium burning favors a rapid loss of the external mantle, which might prevent the star from experiencing the thermal pulses. The high-mass threshold ($8~M_\odot$) is sensitive to the assumption of core overshoot during the main sequence since the core mass at the beginning of the AGB is correlated with the amount of extra-mixing assumed during the core H-burning phase.

The high-mass threshold is valid in this context for the solar and slightly sub-solar metallicity, and it was determined on the basis of the stellar models used in the present investigation, that adopt a moderate overshoot from the external border of the convective core during the main sequence. If we do not take into account extra-mixing, this high-mass threshold would shift to $\sim 10~M_{\odot}$. The high-mass threshold of AGB evolution also depends on metallicity. Since a star develops a more massive core during the main sequence the lower the metallicity, 
the low-mass threshold for carbon ignition is $8~M_{\odot}$ at solar metallicity, but would be $7.5~M_{\odot}$ at  $Z=3\times 10^{-4}$, and even lower for lower metallicities (see e.g. Dell'Agli et al. 2019).

The AGB evolution is characterized by the gradual expansion and cooling of the external regions of the star, which favor the loss of the entire envelope with high rates of mass-loss, and the injection in the interstellar medium of gas reprocessed by internal nucleosynthesis \citep[see][for an exhaustive review]{Herwig2005}. 

The surface chemistry during the AGB phase varies due to two main physical processes, whose relative importance depends on the mass of the progenitor.
Stars with M$<$4M$_{\odot}$ experience repeated episodes of TDU. These are deep inward penetrations of the surface convection during which the innermost layers of the stellar envelope reach triple-$\alpha$ nucleosynthesis sites; such sites are greatly enriched in $^{12}$C, which is then rapidly transported to the stellar surface, owing to the high efficiency of the convective currents \citep[e.g.][]{iben83,busso99}. 

Repeated TDU events can lead the carbon-to-oxygen number ratio to exceed unity (C/O$>$1), and the AGB becomes a carbon star. Stars with M$\geq$4M$_{\odot}$ experience HBB, whose ignition occurs when the temperature of the convective envelope reaches values higher than 30-40 MK, which allows for efficient H-burning via proton capture nucleosynthesis in the most internal regions of the envelope \citep{renzini81,blocker91,sackmann91}. 
This process has the main result of converting carbon into nitrogen in the surface regions, thus preventing the formation of carbon stars and enhancing nitrogen abundances. 

The latest generation of AGB models \citep[see][for a summary review]{karakas14} are the best to describe the detailed evolution of the chemical variation at the stellar surface. By comparing the chemical pattern measured in the PNe with the chemical abundances predicted for the final stage of the AGB evolution for stars with different masses and metallicities, it is possible to characterize the individual sources in terms of their epoch of formation and initial chemistry \citep{VSD15,VSD16,VSD17}. 

For model comparison we use the ATON models \citep{ventura98} at solar \citep[Z=0.014;][]{ventura18} and sub-solar (Z=0.008, 0.004; \citep{ventura13}) metallicities.
These models are presently the only ones where the full integration of the equations of stellar structure and the AGB evolution of the star are self-consistently coupled with the dust formation process in the wind. The models used here have been 
extensively compared with those from other research groups, particularly at solar \citep{ventura18}
and sub-solar \citep{ventura15b, ventura16b} metallicities. 
These comparisons outlined significant dissimilarities in the $M>3~M_{\odot}$ domain for what attains the evolution of the main physical parameters and the modification of the surface chemistry of the stars experiencing HBB, that are related to the differences in the description of turbulent convection, particularly in the inner regions of the convective envelope. On the other hand, consistency was found in the low-mass domain, where stars do not experience the HBB. All sources analyzed in this paper descend from $M\leq 2.5~M_{\odot}$ progenitors; therefore the conclusions
drawn in the present context are substantially independent of the stellar models used.

In Figure~\ref{fig:CNCO} we examine the PN carbon abundances in the context of stellar and nebular evolution. In the left panel, we compare data and models in the (C/H) -- (N/H) plane, and in the right panel in the (C/H) -- (O/H) plane. The plotted data, in black symbols, refer to the
ionic abundances derived in this paper. The errorbars showed in the figure give the typical carbon abundance uncertainty, in dex, if we assume a 5 or 10$\%$ uncertainties in the electron temperatures. Note that uncertainties in electron densities, line fluxes, and reddening are too small to make a difference in the plotted bar, as described in the Analysis section. Also note that these uncertainties stem from an initial guess, and are narrowly distributed, thus can be used for all the plotted points. The model surface chemistry of different initial mass and metallicity are indicated with color symbols. The models with the same initial metallicity have been connected with lines, and the initial stellar masses are also indicated in the Figure.

The ATON models and the chemical loci in Fig.~\ref{fig:CNCO} do not include the effects of deep mixing during the RGB evolution, which is effective for M$<2$M$_{\odot}$. For these models, the N/H in Fig.~\ref{fig:CNCO} is the lower limit of the theoretical expectations for stars exposed to deeper mixing while ascending the RGB. The extra-mixing affects carbon abundances as well in stars with M$\sim 1$M$_{\odot}$. In these stars -- experiencing only a few, if any, TDU events -- the extra-mixing has the effect of lowering C/H, thus the model carbon abundances should be interpreted as upper limits. 

In the following discussion, we match the data points and models, assuming that binary interaction with stellar companions is not affecting the evolution and nucleosynthesis of the primary AGB star. This means we assume that the progenitor stars do not evolve through the common-envelope (CE) stage, i.e., they are either single stars or members of wide binary stars. We found that PNe with carbon abundances can be sorted into two major groups.

\subsection{PNe with low carbon abundances}

Planetary nebulae PN~G025.3-04.6 and PN~G042.9-06.9 (open squares in Fig.~\ref{fig:CNCO}), PN~G053.3+24.0 (crossed square), and PN~G295.3-09.3 (filled square) are characterized by similar, low carbon abundances (log(C/H)+12$\sim$7.9) and similar C/O ratios below unity. The carbon abundance of PN~G053.3+24.0 is uncertain. By their carbon abundances, it is unlikely that any of these PNe had progenitors with mass in the 1.5\,M$_{\odot}\leq$\,M\,$\leq3$\,M$_{\odot}$ range, since, if that was the case, they would exhibit significantly higher carbon abundances (log(C/H)+12$>$8.5). Furthermore, their nitrogen abundances (log(N/H)+12$<$8) seem to indicate that their progenitors did not go through the HBB process, seemingly excluding high-mass ($>$3\,M$_{\odot}$) progenitors. This scenario is reinforced by their low He abundances (log(He/H)+12$<$11.2, see Table \ref{tab:Parameters}), which seem to indicate that their progenitors did not experience a second dredge up.

From the comparison with models (Fig.~\ref{fig:CNCO}, left panel) the carbon abundances of these PNe would be compatible with those expected in the external layers of $\sim$1\,M$_{\odot}$ AGB stars with initial half-solar metallicity (red triangles), but the nitrogen abundances measured for these PNe are higher than the final surface abundances of AGB stars with such metallicity and mass.

The effect of extra-mixing on the RGB is included in the left plot of Figure~\ref{fig:CNCO}. To this end, we estimate the differences of the final yields if we include extra-mixing, for initial masses of 1 and 1.5\,M$_{\odot}$, following the prescriptions of \citet{Lagarde19}, and references therein. The yellow area of the figure indicates the final surface chemical abundances for progenitors in the 1-1.5 M$_{\odot}$ mass range and half-solar metallicity that had experienced extra-mixing on the RGB. If we assume that the progenitor mass of the observed low-carbon PNe is in the $\sim1.1-1.2$\,M$_{\odot}$ range, the extra-mixing would make both the carbon and nitrogen abundances compatible with the observations. 

Most of the PNe in this group have oxygen abundances compatible with roughly half-solar (Z=0.008; open squares inig.~\ref{fig:CNCO}) metallicity models; the only exception is PN~G295.3-09.3 (filled square), which has a lower O abundance. This PN could still derive from a similar evolutionary path to the other PNe in this group, except with a lower metallicity progenitor. The right panel of Fig.~\ref{fig:CNCO} shows well the effect of initial metallicity on the O/H abundances and it is used to resolve the degeneracy between initial composition and CNO evolutionary effects.

Three of the low-carbon PNe (i.e., all except PN~G053.3$+$24.0) have been observed with Spitzer/IRS, thus their dust type is known. All three are ORD (oxygen-rich dust) PNe with amorphous dust type --  PN~G042.9$-$06.9 displays additional weak crystalline silicate features -- in agreement with the gas-phase carbon abundances and the observed nebular C/O ratios below unity, thus reinforcing the connection between the gas-phase to dust-phase chemistry, and our initial-mass and metallicity interpretation.

PNe in the low-carbon group are characterized by morphology that departs from symmetry, such as bipolar and point-symmetric. Observational analysis of large PN samples associates asymmetric morphology with high nitrogen abundances and low Galactic latitude, both hinting to younger, more massive progenitors \citep[e.g.,][]{MVS00}. Interestingly, PNe with low carbon abundance seem to be located away from the Galactic plane (see Table \ref{tab:Parameters}), based on their distances and uncertainties calibrated with Gaia parallaxes \citep{SBL20}, which is incompatible with high mass progenitors. 

From the viewpoint of modeling, bipolar PN morphology has been linked to the presence of binary (sub)stellar companions \citep[e.g.,][]{JO17,DE20}, or to magnetic fields \citep[e.g.,][]{GS97}, 
although it has been shown that strong deviations from spherical symmetry, via the action of magnetic fields, generally require a binary companion \citep[e.g.,][]{NO07,GS14}. It appears that the observations for the low-carbon PNe agree with a low-mass progenitor, possibly with a sub-stellar companion.

\subsection{PNe with enhanced carbon abundances} 

Planetary nebulae PN~G006.1+08.3, PN~G278.6-06.7, PN~G281.0-05.6 (open circles in Fig.~\ref{fig:CNCO}), PN~G107.4-02.6, PN~G275.3-04.7, and PN~G286.0-06.5 (filled circles), and
PN~G068.7+14.8 (not in the Figure for lack of ancillary abundances)
have enhanced C abundance – or lower limit thereof – (log(C/H)+12$\geq$8.5), which is compatible with several TDU episodes in the progenitor star. Following this interpretation, these PNe should have progenitors with masses in the 1.5$-$3.0\,M$_{\odot}$ range, which were formed around 0.25-1.5\,Gyr ago. 
To verify this interpretation we should also consider the N and O abundances available in the literature (see Table \ref{tab:Parameters}). Two nebulae, PN~G006.1+08.3 and PN~G281.0-05.6, have nitrogen and oxygen abundances compatible with masses in the 1.5$-$3.0\,M$_{\odot}$ range, formed with a half-solar metallicity -- or slightly higher. A progenitor of 1.5$-$3.0\,M$_{\odot}$ and metallicity between half-solar and solar is compatible also with PN~G278.6-06.7 (note that above 1.5 M$_{\odot}$ the effects of extra-mixing are marginal, Lagarade et al. 2019). For this PN, a progenitor of a higher mass ($\sim$3.5M$_{\odot}$) and lower metallicity would also comply with the observed C, N, and O abundances, although its argon abundance would rule it out: the expected argon abundance at Z=0.004 is 12+log(Ar/H)=5.65 while the measured abundance is definitively higher \citep[measured log(Ar/H)+12$\sim$6]{GHG14}. The N and O abundances of PN~G286.0-06.5 and PN~G275.3-04.7 are compatible with Z=0.004 models of masses in the range of 1.5$-$3.0\,M$_{\odot}$. The same type of progenitor is plausible for PN~G107.4-02.6, even if in this case N abundance is not available from the literature.

Three of these PNe have round or elliptical morphology, two of them have uncertain morphology from the 2d UV spectrograms (no resolved optical imaging available), and one (PN~G286.0-06.5) is an elongated bipolar. 
All PNe in the high-carbon group have CRD, either aromatic or aliphatic (two objects display both dust types), consistently with the gas-phase abundances, showing complete agreement between the dust-phase and gas-phase carbon.

\section{Discussion}

We determined that several PNe in our sample have C/O$<$1. We can infer that their progenitors did not go through the carbon star phase by comparing their chemistry to the stellar AGB models. PN~G25.3--4.5, PN~G42.9--06.9, and PN~G295.3--09.3
seem to have evolved from $\sim 1.1-1.2$M$_{\odot}$ progenitor stars, whose surface C and N abundances are the results of a few TDU events and deep mixing during the RGB, respectively (yellow area in the left panel of Fig.~\ref{fig:CNCO}). All low-carbon PNe have faint spectra, and they are far from the Galactic plane. The latter observable is consistent with the scenario that the progenitors of these low-carbon PNe are low-mass AGB that were in binary systems, with a sub-stellar body as a companion.  \citet{DE20} has shown that all fourteen C/O$<$1 AGB stars observed with ALMA under their ATOMIUM program are aspherical, suggesting that binary interaction may dominate the evolution of low-mass AGB stars with low C/O. 
It is worth noting that the carbon abundance of PN~G053.3+24.0 is uncertain, thus its classification within this evolutionary group also uncertain. Note also that its optical morphology \citep{SSV16} is rather different from that of the other PNe in this group. 
The group of PNe that have enhanced carbon abundances could have progenitors with masses in the $\sim 1.5-2.5$ M$_{\odot}$ range, which were formed around 0.25-1.5\,Gyr ago. Unfortunately, all carbon abundances of this group of PNe are either lower limits or are uncertain. Nonetheless, their status of carbon-rich PNe is supported by their Spitzer/IRS CRD dust types. 

In Figure 6 we plot the log(C/O) vs. log(O/H)+12 for the compact Galactic PNe studied here and elsewhere, together with the samples of Magellanic Cloud PNe. In this plot we included all compact PNe with UV-based carbon abundances published in the literature or studied in this paper. We indicate the PN population by the symbol shape: triangles for the SMC, squares for the LMC, circles for compact Galactic PN (this study;  O, N, and C abundances from Tables 2 and 9); plus signs, crosses, and asterisks for compact Galactic PNe from other studies, respectively, \citet{2015ApJ...803...23D, 2000ApJ...531..928H, KB94}. We use the symbol's color to indicate the dust status of the PNe from Spitzer/IRS spectroscopy: Cyan symbols: featureless dust spectra, Red symbols: carbon-rich dust (CRD) PNe; blue symbols: oxygen-rich dust (ORD) PNe, black symbols: no IRS dust information. Since we selected the Galactic PNe, both from this study and from the literature, based on their apparent sizes ($\theta<$5\arcsec), their diameters are smaller than the Spitzer/IRS aperture, thus the comparison with Magellanic Cloud and compact Galactic PNe of the other samples is meaningful, as all spectra include the flux from the whole nebular surface.
We found complete segregation of CRD PNe in the C/O$>$1 quadrant, and of ORD PNe in the C/O$<$1 quadrant. This occurs independently on stellar or galactic metallicity. Our carbon analysis indicates that the sample studied here has predominantly super solar carbon abundance, with median carbon $<$C/H$>_{\rm med}\pm\sigma=6.31\pm3.52\times10^{-4}$.
We plot on the Figure the fit by \citet{Nicholls} derived by interpolating stellar abundances (see references cited therein). The nebular enrichment of carbon is clearly seen for CRD PNe, independent of the studied population, a confirmation of PN carbon enrichment role \citep[e.g.]{2018MNRAS.473..241H}, with the added value of the correlation with dust composition. 
It is worth noting that the correspondence between dust and gas abundances -- i.e., all CRD PNe have C/O$>$1, and all ORD PNe have C/O$<$1 -- is stronger in our study than in the work by \citet{DIR14}, who found a few exceptions to this correspondence, likely due to the mismatch between the {\it Spitzer} and other spectral apertures, and the inclusion of extended Galactic PNe in their sample.

\section{Summary}
We selected 75 compact, or moderately extended, Galactic PNe to be observed with {\it HST}/STIS through the G230L and G140L gratings to detect their UV emission lines for carbon abundance measurements. Only 30 of the targets have been observed in two ''snapshot" programs, and we measured carbon abundances of 11 targets. With the support of ancillary data sets we found a striking correlation between gas-phase (this and other studies of UV-based carbon abundances in compact Galactic PNe) and dust-phase (Spitzer/IRS) carbon abundances, i.e., {\it all} carbon-rich dust (CRD) PNe studied here have C/O$>$1, and all ORD PNe have C/O$<$1. By studying these correlations together with those found in Magellanic Cloud PNe we found that this one-to-one correlation is independent of the initial progenitor's metallicity. 
We compared the loci of the C, N, O abundance patterns on different diagnostic planes for our PN sample with the footprints of the final yields from stellar evolution models. We found that the progenitors of most carbon-poor PNe are likely in the M/M$_{\odot}\le 1.2$ range, with slightly sub-solar metallicity. Identifying such old progenitors is
 useful to calibrate a radial metallicity gradient for old Galactic probes \citep{SH18}. It is worth noting that, while Gaia distances from parallaxes are not available for all the CSs of the compact Galactic PNe studied here, statistical distances based on Gaia DR2 parallaxes indicate that PNe in this group are generally far from the Galactic plane, an additional indication of a very old Galactic population.
We also found that the carbon-enhanced PNe in our sample are the likely progeny of carbon stars in the $1.5\le$ M/M$_{\odot}\le 3$ range.
This work presents a limited but important sample of carbon abundances from UV lines in compact Galactic PNe, it augments considerably the number of Galactic PNe whose carbon abundances have been measured based on {\it HST} spectra, and it greatly expands the sample for which gas-phase and dust-phase carbon can be simultaneously available in compact PNe.

\acknowledgments

This research is based on observations made with the NASA/ESA Hubble Space Telescope obtained from the Space Telescope Science Institute, which is operated by the Association of Universities for Research in Astronomy, Inc., under NASA contract NAS 5-–26555. These observations are associated with GO programs 15211 and 16013. We thank an anonymous Referee for important comments on this paper. We acknowledge the usage of the pyneb package, and thank Christophe Morisset for his help. DAGH acknowledges support from the ACIISI, Gobierno de Canarias and the European Regional Development Fund (ERDF) under grant with reference PROID2020010051 as well as from the State Research Agency (AEI) of the Spanish Ministry of Science and Innovation (MICINN) under grant PID2020-115758GB-I00.

\vspace{5mm}
\facilities{HST(STIS)}
\startlongtable
\begin{deluxetable*}{lllllrD}
\tablecaption{Observing Log\label{tab:ObsLog}}
\tablewidth{0pt}
\tablehead{
\colhead{PN G} & \colhead{Name} & \colhead{Date} & \colhead{Obs ID\tablenotemark{a}} & \colhead{Grating} & \colhead{Duration\tablenotemark{b}} & \multicolumn2c{Aperture}\\
& & & & & \colhead{(s)} & \multicolumn2c{(arcsec)}
}
\decimals
\startdata
003.9--14.9 & Hb 7 & 2019 May 22 & odk350kqq & G140L & 218 & 2.0 \\
 & & & odk350krq & G230L & 185 & {} \\
004.3--02.6 & H 1-53 & 2019 May 23 & odk370tzq & G140L & 218 & 0.25 \\
 & & & odk370u0q & G230L & 185 & {} \\
006.1+08.3 & M 1-20 & 2018 Jun 06 & odk362khq & G140L & 0 & . \nodata \\
 & & & odk362kiq & G230L & 1205 & 2.0 \\
011.1+07.0 & Sa 2-237 & 2019 Mar 09 & odk354yeq & G140L & 37 & 0.25 \\
 & & & odk354yfq & G230L & 30 & {} \\
025.3--04.6 & K 4-8 & 2018 Aug 28 & odk304haq & G140L & 236 & 1.00 \\
 & & & odk304hbq & G230L & 208 & {} \\
032.5--03.2 & K 3-20 & 2018 Sep 10 & odk357d9q & G140L & 1200 & 0.25 \\
 & & & odk357daq & G230L & 1200 & {} \\
038.4--03.3 & K 3-20 & 2018 Sep 10 & odk358buq & G140L & 1200 & 0.50 \\
 & & & odk358bvq & G230L & 1200 & {} \\ 
038.7--03.3 & M 1-69 & 2018 May 28 & odk371d1q & G140L & 152 & 0.25 \\
 & & & odk371d2q & G230L & 155 & {} \\
042.9--06.9 & NGC 6807 & 2018 Nov 05 & odk306gaq & G140L & 20 & 1.00 \\
 & & & odk306gbq & G230L & 17 & {} \\
048.5+04.2 & K 4-16 & 2018 Nov 05 & odk308ieq & G140L & 1200 & 0.25 \\
 & & & odk308ifq & G230L & 1200 & {} \\
053.3+24.0 & Vy 1-2 & 2019 Mar 30 & odk310xcq & G140L & 206 & 3.00 \\
 & & & odk310xdq & G230L & 169 & {} \\
068.7+14.8 & Sp 4-1 & 2018 Aug 06 & odk313a3q & G140L & 155 & 1.25 \\
 & & & odk313a4q & G230L & 180 & {} \\
095.2+00.7 & K 3-62 & 2018 Jul 22 & odk364pdq & G140L & 1200 & 0.25 \\
 & & & odk364peq & G230L & 1200 & {} \\
097.6--02.4 & M 2-50 & 2019 Jul 11 & odk315o2q & G140L & 1200 & 0.275 \\
 & & & odk315o3q & G230L & 1200 & {} \\
107.4--02.6 & K 3-87 & 2018 Mar 06 & odk318n3q & G140L & 1200 & 0.275 \\
 & & & odk318n4q & G230L & 1200 & {} \\
232.8--04.7 & M 1-11 & 2018 Jun 02 & odk367bwq & G140L & 1200 & 1.50 \\
 & & & odk367bxq & G230L & 1200 & {} \\
264.4--12.7 & He 2-5 & 2018 Aug 20 & odk321egq & G140L & 175 & 2.75 \\
 & & & odk321ehq & G230L & 149 & {} \\
275.3--04.7 & He 2-21 & 2020 Jul 28 & oe7322hvq & G140L & 1200 & 2.50 \\
& & & oe7322hwq & G230L & 1200 & {} \\
278.6--06.7& He 2-26 & 2018 Jan 07 & odk323ruq & G140L & 175 & 2.50 \\
 & & & odk323rvq & G230L & 154 & {} \\
281.0--05.6 & IC 2501 & 2018 Mar 03 & odk369fcq & G140L & 30 & 5.00 \\
 & & & odk369fdq & G230L & 26 & {} \\
285.4+01.5 & Pe 1-1 & 2019 Jun 10 & odk324c6q & G140L & 1200 & 0.25 \\
 & & & odk324c7q & G230L & 1200 & {} \\
285.4+02.2 & Pe 2-7 & 2019 Feb 02 & odk325ngq & G140L & 1200 & 0.25 \\
 & & & odk325nhq & G230L & 1200 & {} \\
%
286.0--06.5 & He 2-41 & 2019 Feb 02 & odk326ydq & G140L & 623 & 2.00 \\
 & & & odk326yeq & G230L & 567 & {} \\
 & & 2019 Oct 31 & oe7326rmq & G140L & 623 & {} \\
 & & & oe7326rnq & G230L & 567 & {} \\
295.3--09.3 & He 2-62 & 2018 Jan 05 & odk329d3q & G140L & 87 & 1.00 \\
 & & & odk329d4q & G230L & 76 & {} \\
309.0+00.8 & He 2-96 & 2018 Jun 15 & odk331e1q & G140L & 1200 & 0.25 \\
 & & & odk331e2q & G230L & 1200 & {} \\
336.9+08.3 &  St Wr 4-10& 2019 Feb 27 & odk337pdq & G140L & 0 & . \nodata \\
 & & & odk337peq & G230L & 0 &  \nodata \\
340.9--04.6 & Sa 1-5 & 2020 Aug 27 & oe7338stq & G140L & 1200 & 1.25 \\
& & & oe7338suq & G230L & 1200 & {} \\
343.4+11.9 &  H 1-1& 2020 Sep 20 & oe7340fxq & G140L & 0 & \nodata \\
& & & oe7340fyq & G230L & 0 & {} \\ 
351.3+07.6 & H 1-4 & 2019 May 05 & odk345h5q & G140L & 842 & 1.00 \\
 & & & odk345h6q & G230L & 765 & {} \\
355.2--02.5 & H 1-29 & 2019 Jun 17 & odk374feq & G140L & 496 & 1.00 \\
 & & & odk374ffq & G230L & 512 & {} \\ 
\enddata
\tablenotetext{a}{Observation identifiers beginning with ``odk3'' correspond to {\it HST} program GO-15211, and those beginning with ``oe73'' correspond to GO-16013.}
\tablenotetext{b}{A duration of zero indicates an on-board failure of the exposure.}
\end{deluxetable*}

\clearpage

\begin{deluxetable}{ l l r  r r l l }    
\tablewidth{0pt}
\tabletypesize{\footnotesize{}}
		\tablecolumns{5}
\tablewidth{0pt}
\tablecaption{PN parameters\label{tab:Parameters}}
\tablehead {\colhead{PN~G}& \colhead{$|z_{\rm kpc}|$\tablenotemark{a}}& 
\colhead{${\rm log}(He/H)+12$\tablenotemark{b}}& \colhead{${\rm log}(N/H)+12$\tablenotemark{b}}& \colhead{${\rm log}(O/H)+12$\tablenotemark{b}}& \colhead{Morph.}& \colhead{Dust Type\tablenotemark{c}} \\}
\startdata 
006.1+08.3  &   1.09& 
11.02$\pm$0.05&  7.78$\pm$0.08& 8.56$\pm$0.08&  E\tablenotemark{d}&   CRD; aromatic/aliphatic\\
025.3--04.6  &  0.84&  
11.01$\pm$0.03&  7.79$\pm$0.06& 8.59$\pm$0.09&  P\tablenotemark{e}&    ORD; amorphous\\
042.9--06.9\tablenotemark{f}&  0.84&     
10.99$\pm$0.03&  8.00$\pm$0.24& 8.57$\pm$0.08&B\tablenotemark{e}&   ORD; crystalline/amorphous\\
053.3+24.0\tablenotemark{f}&   3.16&    
11.23$\pm$0.05&  7.90$\pm$0.05& 8.46$\pm$0.05&   B\tablenotemark{e}&   N/A\\
068.7+14.8&  3.24&    
$\dots$& $\dots$& $\dots$&   R\tablenotemark{e}&   CRD; aromatic\\
107.4--02.6&  0.56&    
11.01$\pm$0.05&  $\dots$& 8.29$\pm$0.19&  E\tablenotemark{e}&   CRD; aliphatic\\
275.3--04.7&  0.37&    
11.08$\pm$0.04& 7.64$\pm$0.09& 8.44$\pm$0.10&  E\tablenotemark{e}&   CRD; aliphatic\\
278.6--06.7&  0.45&    
11.00$\pm$0.04&  8.00$\pm$0.13& 8.52$\pm$0.11&  E\tablenotemark{e}&   CRD; aliphatic\\  
281.0--05.6& 0.61&     
$\dots$& 8.16$\pm$0.00&  8.63$\pm$0.10&   E\tablenotemark{e}&   CRD; aromatic/aliphatic\\
286.0--06.5&  0.76&     
10.99$\pm$0.03&  7.66$\pm$0.08& 8.32$\pm$0.07&   B\tablenotemark{e}&   CRD; aliphatic\\
295.3--09.3& 1.63&     
11.01$\pm$0.03&  7.87$\pm$0.12&  8.15$\pm$0.07&   B\tablenotemark{e}&   ORD; amorphous\\
   351.3+07.6& 2.90&     
   $\dots$& $\dots$& $\dots$&    BC\tablenotemark{e}&   ORD; amorphous\\
    \enddata
\tablenotetext{a}{Calculated using the Gaia DR2-calibrated distance scale \citep{SBL20}, where d$_{\rm scale}$/d$_{\rm par}$=0.95$\pm$0.25. PN~G251.3+07.6 distance was derived directly from the DR2 Gaia parallax.}
\tablenotetext{b}{The He, N, and O abundances used in this study are from \citet{GHG14} except for PN~G042.9--06.9 and PN~G275.3-04.7 \citep{PMS}, and PN~G053.3+24.0 \citep{S06}.}
\tablenotetext{c}{Dust types from \citet{SGG12}, except for PN~G006.1+08.3 \citep{PC09,GH10} and PN~G281.0--05.6 \citep{OT14}.}
\tablenotetext{d}{Uncertain morphology, based only on UV slitless spectroscopy (this study).}
\tablenotetext{e}{Morphology derived from WFC3 imaging through a selection of filters \citep{SSV16}. }
\tablenotetext{f}{May be a halo PN.}
\end{deluxetable}

\pagebreak
\begin{deluxetable*}{lc DD h DD h DD}
\tablecaption{Relative Emission Line Fluxes\label{tab:Flux1}}
\tablewidth{0pt}
\tabletypesize{\footnotesize{}}
\tablehead{
\colhead{Wave}  & \colhead{ID} & 
\multicolumn4c{006.1+08.3} & \colhead{} & 
\multicolumn4c{025.3--04.6}  & \colhead{} & 
\multicolumn4c{042.9--06.9}\\
\cline{3-6} \cline{8-11} \cline{13-16} 
\colhead{(\AA)} &  & 
\multicolumn2c{F$_{\lambda}$} & \multicolumn2c{I$_{\lambda}$} & & 
\multicolumn2c{F$_{\lambda}$} & \multicolumn2c{I$_{\lambda}$} & & 
\multicolumn2c{F$_{\lambda}$} & \multicolumn2c{I$_{\lambda}$} 
}
\decimals
\startdata
1907+09 & \ion{C}{3}] & 
    5.08\pm0.46 & 173.92\pm15.75 & & 
    10.03\pm0.96 & 45.65\pm4.39 & &
    6.86\pm0.51 & 23.53\pm1.76 \\    
$\sim$2810 & ? & 
    4.57\pm0.13 & 26.55\pm0.74 & & 
        . \nodata & . \nodata & & 
    . \nodata & . \nodata \\
\hline
  & log F$_{H\beta}$ & 
    \multicolumn{4}{c}{$-11.93\pm0.01$} & & 
    \multicolumn{4}{c}{$-12.44\pm0.10$} & &
        \multicolumn{4}{c}{$-11.41\pm0.01$} \\
  & $c_{H\beta}$ & 
    \multicolumn{4}{c}{$1.17\pm0.10$} & & 
    \multicolumn{4}{c}{$0.50\pm0.10$} & &
      \multicolumn{4}{c}{$0.41\pm0.10$} \\
\enddata
\end{deluxetable*}

\begin{deluxetable*}{lc DD h DD h DD}
\tablecaption{Relative Emission Line Fluxes --- Continued\label{tab:Flux2}}
\tablewidth{0pt}
\tabletypesize{\footnotesize{}}
\tablehead{
\colhead{Wave}  & \colhead{ID} & 
\multicolumn4c{053.3+24.0} & \colhead{} & 
\multicolumn4c{068.7+14.8} & \colhead{} & 
\multicolumn4c{107.4--02.6} \\
\cline{3-6} \cline{8-11} \cline{13-16} 
\colhead{(\AA)} &  & 
\multicolumn2c{F$_{\lambda}$} & \multicolumn2c{I$_{\lambda}$} & & 
\multicolumn2c{F$_{\lambda}$} & \multicolumn2c{I$_{\lambda}$} & & 
\multicolumn2c{F$_{\lambda}$} & \multicolumn2c{I$_{\lambda}$} \\
}
\decimals
\startdata
1640 & \ion{He}{2} & 
    88.67\pm0.42 & 101.63\pm0.48 & & 
     . \nodata & . \nodata & & 
    . \nodata & . \nodata \\
1907+09 & \ion{C}{3}] & 
    15.39\pm0.72 & 17.90\pm0.84 & & 
    197.86\pm0.67 & 615.87\pm2.08 & &
        9.10\pm0.34 & 396.6\pm14.86 \\
2325--29 & \ion{C}{2}] & 
    20.74\pm0.52 & 24.11\pm0.60 & & 
    38.24\pm0.70 & 118.48\pm2.15 &&
     . \nodata & . \nodata \\
2470 & [\ion{O}{2}] & 
    3.11\pm0.31 & 3.49\pm0.34 & & 
     . \nodata & . \nodata & & 
    . \nodata & . \nodata \\[0.1cm]
3023 & \ion{O}{3} & 
    4.89\pm0.52 & 5.20\pm0.55 & & 
     . \nodata & . \nodata & & 
    . \nodata & . \nodata \\
3043+47 & \ion{O}{3} & 
    2.09\pm0.55 & 2.22\pm0.59 & &
     . \nodata & . \nodata & & 
    . \nodata & . \nodata \\
\hline
  & log F$_{H\beta}$ & 
    \multicolumn{4}{c}{$-11.51\pm0.01$} & & 
    \multicolumn{4}{c}{$-11.95\pm0.10$} &&
    \multicolumn{4}{c}{$-13.21\pm0.20$} \\ 
  & $c_{H\beta}$ & 
    \multicolumn{4}{c}{$0.05\pm0.10$} & & 
    \multicolumn{4}{c}{$0.38\pm0.10$} &&
     \multicolumn{4}{c}{$1.25\pm0.10$} \\ 
\enddata
\end{deluxetable*}
\pagebreak

\begin{deluxetable*}{lc DD h DD h DD}
\tablecaption{Relative Emission Line Fluxes\label{tab:Flux3}}
\tablewidth{0pt}
\tabletypesize{\footnotesize{}}
\tablehead{
\colhead{Wave}  & \colhead{ID} & 
\multicolumn4c{275.3--04.7} & \colhead{} &
\multicolumn4c{278.6--06.7} & \colhead{} & 
\multicolumn4c{281.0--05.6} \\
\cline{3-6} \cline{8-11} \cline{13-16} 
\colhead{(\AA)} &  & 
\multicolumn2c{F$_{\lambda}$} & \multicolumn2c{I$_{\lambda}$} & & 
\multicolumn2c{F$_{\lambda}$} & \multicolumn2c{I$_{\lambda}$} & & 
\multicolumn2c{F$_{\lambda}$} & \multicolumn2c{I$_{\lambda}$} \\
}
\decimals
\startdata
1548+50 & \ion{C}{4} & 
   134.6\pm0.44 & 1206.5\pm3.94 & &
   33.89\pm0.28 & 141.51\pm1.19 & &
     . \nodata & . \nodata \\
1640 & \ion{He}{2} & 
    48.97\pm0.42 &  396.5\pm3.40 & &
    18.70\pm0.22 & 73.03\pm0.87 & &
      . \nodata & . \nodata \\
1907+09 & \ion{C}{3}] & 
    120.1\pm1.28 & 1139.9\pm12.15 & &
    137.67\pm0.32 & 621.27\pm1.46 & &
       54.73\pm0.47 & 271.20\pm2.32 \\ 
2325 & \ion{C}{2}] &
    . \nodata &  . \nodata & &
        16.71\pm3.31 & 74.97\pm1.48 & &
      . \nodata & . \nodata \\
2424 & [\ion{Ne}{4}] &
    11.81\pm0.97 &  88.09\pm7.23 & &
         . \nodata & . \nodata & &
      . \nodata & . \nodata \\
2836 & \ion{O}{3} & 
      . \nodata & . \nodata & & 
    3.53\pm0.18 & 7.32\pm0.37 & &
          . \nodata & . \nodata \\
3133 & \ion{O}{3} & 
      . \nodata & . \nodata & & 
    6.39\pm0.53 & 11.31\pm0.94 & &
          . \nodata & . \nodata \\
\hline
  & log F$_{H\beta}$ & 
     \multicolumn{4}{c}{$-12.15\pm0.10$} & &
    \multicolumn{4}{c}{$-11.55\pm0.01$} & &
       \multicolumn{4}{c}{$-10.67\pm0.01$} \\
  & $c_{H\beta}$ & 
    \multicolumn{4}{c}{$0.8034\pm0.10$} & &    
    \multicolumn{4}{c}{$0.50\pm0.10$} &  &
      \multicolumn{4}{c}{$0.53\pm0.05$} \\ 
\enddata
\end{deluxetable*}

\begin{deluxetable*}{lc DD h DD h DD}
\tablecaption{Relative Emission Line Fluxes\label{tab:Flux4}}
\tablewidth{0pt}
\tabletypesize{\footnotesize{}}
\tablehead{
\colhead{Wave}  & \colhead{ID} & 
\multicolumn4c{286.0--06.5} & \colhead{} & 
\multicolumn4c{295.3--09.3} & \colhead{} & 
\multicolumn4c{351.3+07.6} \\
\cline{3-6} \cline{8-11} \cline{13-16} 
\colhead{(\AA)} &  & 
\multicolumn2c{F$_{\lambda}$} & \multicolumn2c{I$_{\lambda}$} & & 
\multicolumn2c{F$_{\lambda}$} & \multicolumn2c{I$_{\lambda}$} & &
\multicolumn2c{F$_{\lambda}$} & \multicolumn2c{I$_{\lambda}$} 
}
\decimals
\startdata
1640 & \ion{He}{2} & 
    . \nodata & . \nodata & & 
    . \nodata & . \nodata & & 
    8.60\pm0.63 & 56.10\pm4.09 \\
1658--66 & [\ion{O}{3}] & 
    2.00\pm0.12 & 13.35\pm0.79 & & 
    . \nodata & . \nodata & &
    . \nodata & . \nodata \\
1907+09 & \ion{C}{3}] & 
    76.02\pm0.12 & 620.04\pm1.43 & & 
    33.53\pm0.45 & 127.40\pm11.72 & &
    . \nodata & . \nodata \\[0.1cm]
2325--29 & \ion{C}{2}] & 
    10.49\pm0.18 & 84.81\pm1.48 & & 
    . \nodata & . \nodata & & 
    . \nodata & . \nodata \\
2470 & [\ion{O}{2}] & 
    2.91\pm0.18 & 14.15\pm0.85 & & 
    7.02\pm0.60 & 19.21\pm1.64 & &
    . \nodata & . \nodata \\
2836 & \ion{O}{3} & 
    . \nodata & . \nodata & & 
    3.82\pm0.40 & 7.27\pm0.76 & &
    . \nodata & . \nodata \\
\hline
  & log F$_{H\beta}$ & 
    \multicolumn{4}{c}{$-11.90\pm0.10$} & & 
    \multicolumn{4}{c}{$-11.94\pm0.10$} & &
    \multicolumn{4}{c}{$-12.35\pm0.10$} \\
  & $c_{H\beta}$ & 
    \multicolumn{4}{c}{$0.70\pm0.10$} & & 
    \multicolumn{4}{c}{$0.44\pm0.10$} & &
    \multicolumn{4}{c}{$0.69\pm0.10$} \\
\enddata
\end{deluxetable*}

\begin{deluxetable*}{lrllrrl}
\tabletypesize{\scriptsize}    
		\tablewidth{0pt}
		\label{tab:Diag}
		\tablecaption{Plasma diagnostics}            
		\tablehead {
			\colhead{PN G}& \colhead{log(N$_{\rm e})$}& \colhead{Ion}& \colhead{Ref.}&\colhead{T$_{\rm e}$}& \colhead{Ion}& \colhead{Ref.} \\
			\colhead{}& \colhead{[cm$^{-3}$]}&  \colhead{}& \colhead{} & \colhead{[$10^3$ K]}& \colhead{}& \colhead{}\\
		}
\startdata 
%
006.1+08.3    & 4.00 & [\ion{S}{2}] &   WL07  &9.86 & [\ion{O}{3}] & WL07 \\
025.3--04.6 & 4.06& [\ion{S}{2}] & GHG14   & 10.64 & [\ion{O}{3}] &  GHG14 \\
042.9--06.9 &  5.00 & [\ion{S}{2}] &   GHG14    & 10.27 & [\ion{O}{3}] &  GHG14 \\
053.3+24.0  & 3.06 & [\ion{S}{2}] & WLB05 & 10.40   & [\ion{O}{3}] & WLB05 \\
068.7+14.8 & 3.27 & [\ion{O}{2}] & WLB05   & 11.24 & [\ion{O}{3}] & WLB05 \\
107.4--02.6 & 3.04 & [\ion{S}{2}] & K96  &  10.30 &  [\ion{O}{3}] & K96 \\
275.3-04.7& 3.08& [\ion{O}{2}] & GHG14   & 10.11 & [\ion{N}{2}] & GHG14 \\
 & $\dots$ & $\dots$      &      $\dots$   & 12.98 & [\ion{O}{3}] &  GHG14 \\
278.6--06.7 & 3.63 & [\ion{S}{2}] & SSG12    & 11.63 & [\ion{O}{3}] & \tablenotemark{a}\\
281.0--05.6 & 3.62 & [\ion{S}{2}] & \tablenotemark{a} & 9.87 & [\ion{O}{3}] & K86 \\*
286.0--06.5 & 3.35 & [\ion{S}{2}] & SGG12   & 11.12 & [\ion{O}{3}] & \tablenotemark{a}\\
295.3--09.3 & $>$4.00\tablenotemark{b} & [\ion{S}{2}] & \tablenotemark{a}   & 11.87 & [\ion{O}{3}] & \tablenotemark{a} \\
351.3+07.6  & $\dots$ & $\dots$ & $\dots$ & 11.77& [\ion{O}{3}] & \tablenotemark{a} \\
\enddata
\tablerefs{GHG14: \cite{GHG14}; K86: \citet{K86}; SK89: \citet{SK89}; SGG12: \citet{SGG12};  WLB05: \citet{WLB05}; 	WL07: \citet{WL07}}
\tablenotetext{a}{This study; diagnostics flux ratios are from \citet{AMO92}.}
\tablenotetext{b}{For abundance analysis we assume log N$_e = 4.5$}
\end{deluxetable*}

	\begin{deluxetable}{ l ll}    
		\tablecolumns{3}
		\tablewidth{0pt}
		\tablecaption{References, atomic data}            
		\tablehead {
			\colhead{Ion}& \colhead{A-values}& \colhead{Collisional excitation} \\
	}
		\startdata 
		C$^{+}$& \citet{Galavis98}& \citet{Blum92}\\
		C$^{+2}$& \citet{Wiese96}& \citet{Berrington85}\\
		C$^{+3}$& \citet{Wiese96}& \citet{Aggarwal2004}\\
		O$^{+}$& \citet{Zeippen1982}&  \citet{Kisielius2009}\\
		Ne$^{3+}$& \citet{Godefroid1984}& \citet{Giles1981}\\
	\enddata
	\end{deluxetable}

	\begin{deluxetable}{ l r r r r l}    
		\tabletypesize{\footnotesize}
		\tablecolumns{5}
		\tablewidth{0pt}
		\label{tab:Carbon}
		\tablecaption{Carbon abundances}            
		\tablehead {
			\colhead{PN~G}& \colhead{log($C^+/H^+$)}& \colhead{log($C^{2+}/H^+$)} & \colhead{log($C^{3+}/H^+$)}&
			\colhead{ICF(C)}& \colhead{log$(C/H)+12$} \\
	}
		\startdata 
                006.1+08.3 & $\dots$ & -3.42& $\dots$ & $\dots$& 8.58\tablenotemark{a}\\
                025.3--04.6 & $\dots$ & $-4.14$ & $\dots$ & 1.04& 7.88\\
                042.9--06.9 & $\dots$ & -4.29& $\dots$ & 1.90& 7.99\\
                053.3+24.0 & -4.44 & -4.48 & $\dots$&  $\dots$& 7.84\tablenotemark{b} \\
                068.7+14.8 & -3.96 &  -3.19& $\dots$& $\dots$& 8.88\\
                107.4--02.6 &   $\dots$ & -3.10& $\dots$ & $\dots$ & 8.90\tablenotemark{a}\\  
                275.3--04.7 & $\dots$& -3.34& -3.67& 1.04& 8.85\\
                278.6--06.7 & -4.24& -3.29  & -4.22& $\dots$& 8.80 \\
                281.0--05.6 & $\dots$ &  -3.12&  $\dots$& 1.08& 8.92\\
                286.0--06.5 & -4.07 & -3.15 &  $\dots$&  $\dots$& 8.90\\
                295.3--09.3 & $\dots$ & -4.05& $\dots$ & $\dots$& 7.95\tablenotemark{a}\\
%
	\enddata
		\tablenotetext{a}{The atomic abundance is a lower limit because we could not correct for unseen emission lines}
		\tablenotetext{b}{The atomic abundance is uncertain (see text).}
	\end{deluxetable}
	
		\begin{deluxetable}{ l r r  r r}    
		\tabletypesize{\footnotesize}
		\tablecolumns{5}
		\tablewidth{0pt}
		\label{tab:OtherAb}
		\tablecaption{Other abundances}            
		\tablehead {
			\colhead{PN~G}& \colhead{log($He^2+/H^+$)}& \colhead{log($O^{+}/H^+$)} &  \colhead{log($Ne^{3+}/H^+$)}& \colhead{} \\
		}
		\startdata 
                053.3+24.0 & -1.89 & -4.54 & \\
                275.3-04.7 &   $\dots$&  $\dots$&    -4.41\\
                278.6--06.7 & -2.04 &  $\dots$  &  $\dots$ \\
                286.0--06.5 &  $\dots$& -4.00  &  $\dots$ \\
                295.3--09.3 & $\dots$ & -4.23&  $\dots$ \\
                351.3+07.6 & -2.15 & $\dots$ & $\dots$ \\
		\enddata
	\end{deluxetable}

\begin{figure}[ht!]
\plotone{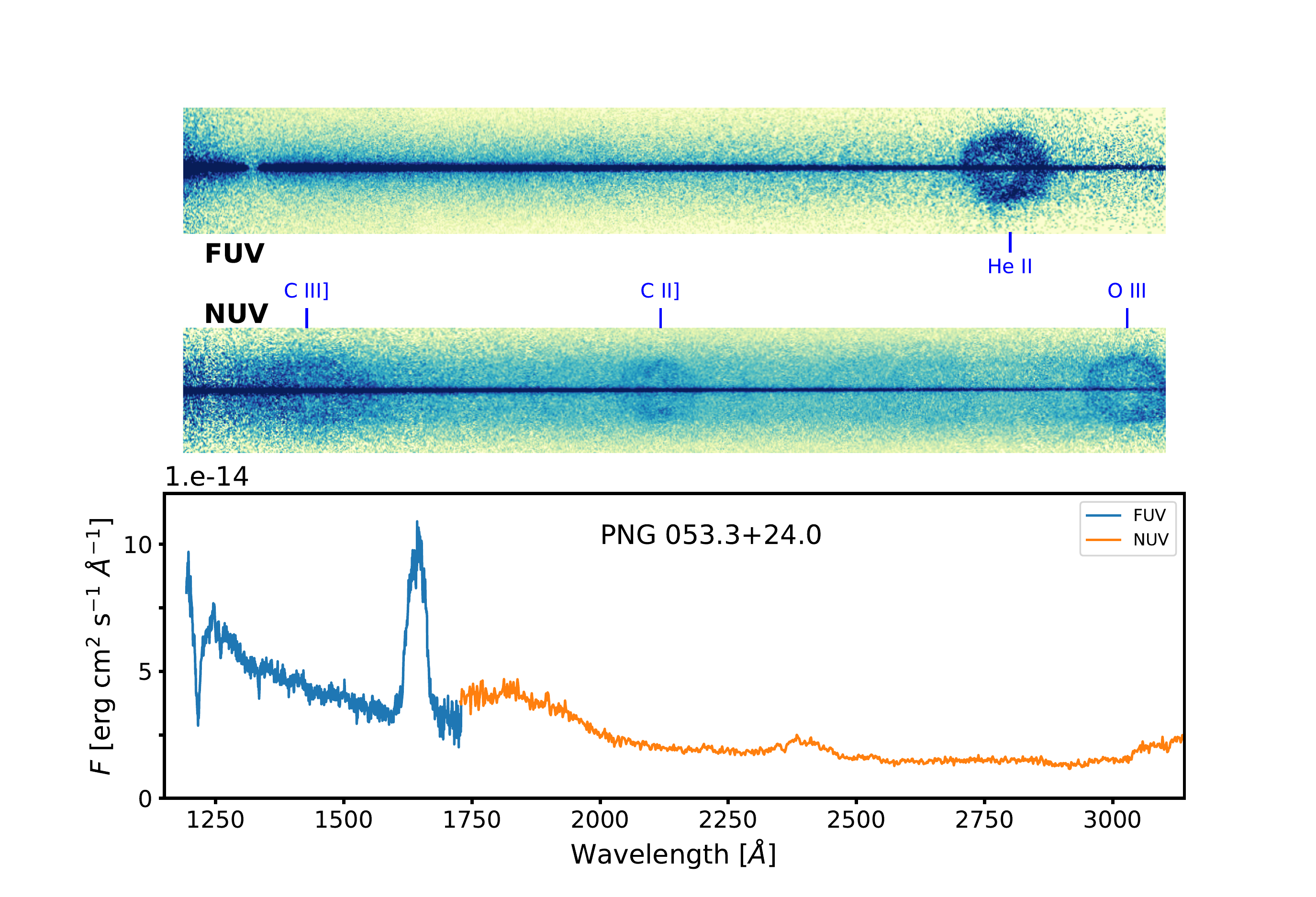}
\caption{False-color rendering of the spectrograms for PN~G053.3+24.0 in the Far-UV (upper) and Near-UV (middle), with identifications of the strongest emission lines (blue labels). The vertical extent corresponds to the boundaries of the extraction aperture. Note that significant NUV emission originates from the superposition of many weak, blended nebular emission lines. Also shown (lower) are the 1-D summed spectra in the Far-UV (blue curve) and Near-UV (orange curve).}

   \end{figure}

\begin{figure}[ht!]

\plotone{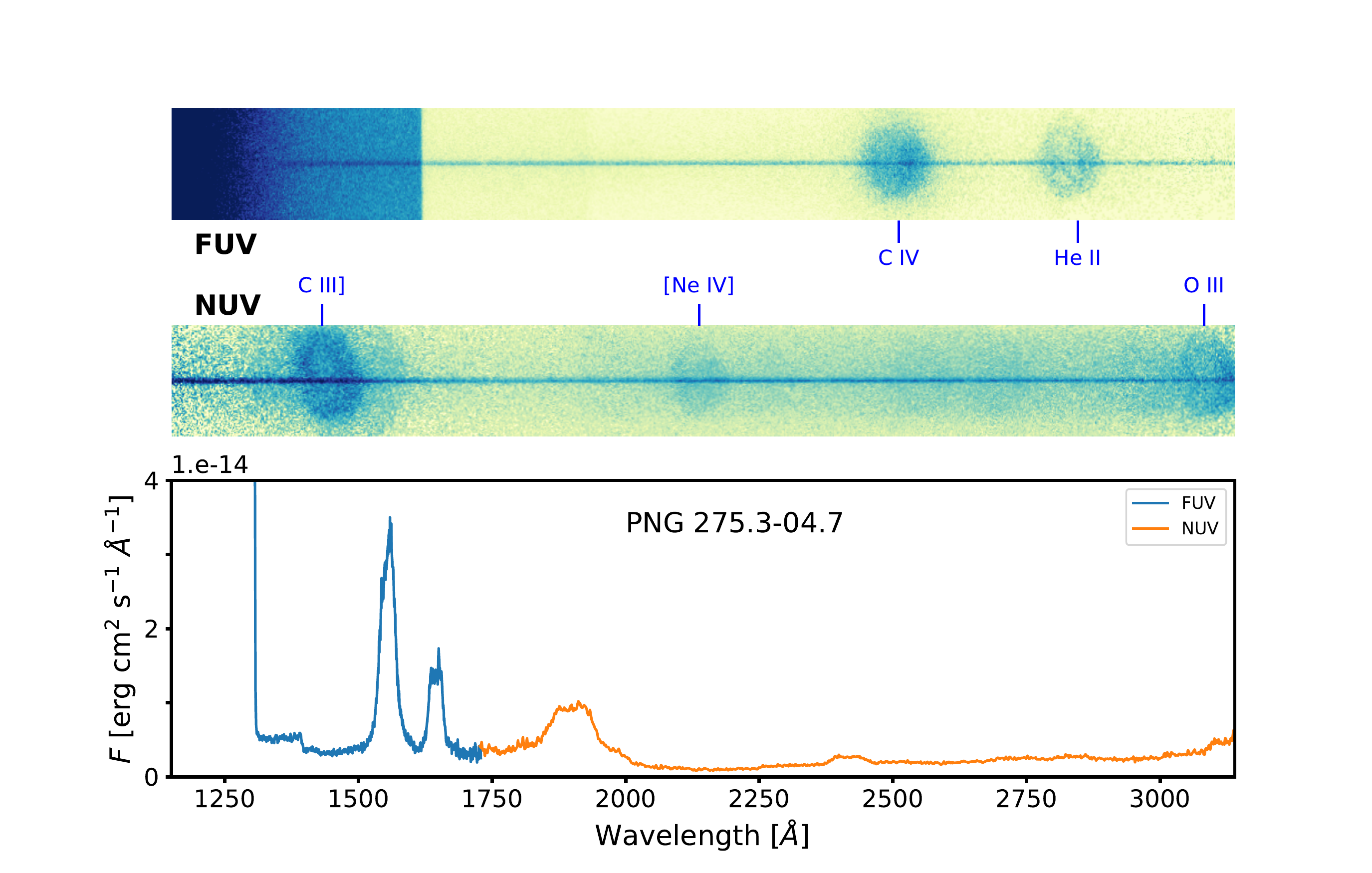}
\caption{False-color rendering of the spectrograms, as in Fig.~1 but for PN~G275.3--04.7. The bright region shortward of $\sim1300$~\AA\ is from geocoronal L$\alpha$.}
\end{figure}
 
\begin{figure}[ht!]
\plotone{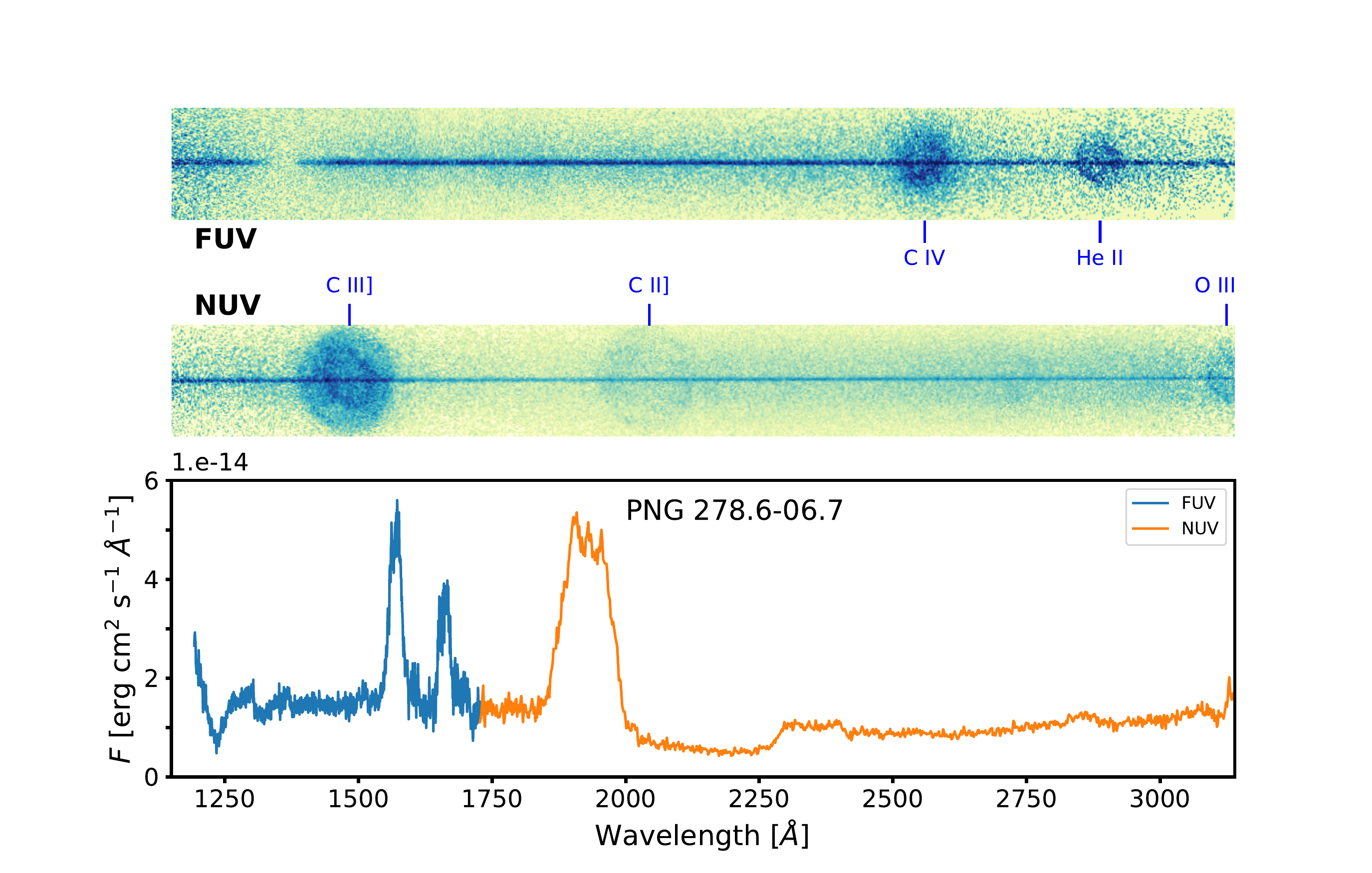}
\caption{False-color rendering of the spectrograms, as in Fig.~1 but for PN~G278.6--06.7.}
\end{figure}

 \begin{figure}[ht!]
\plotone{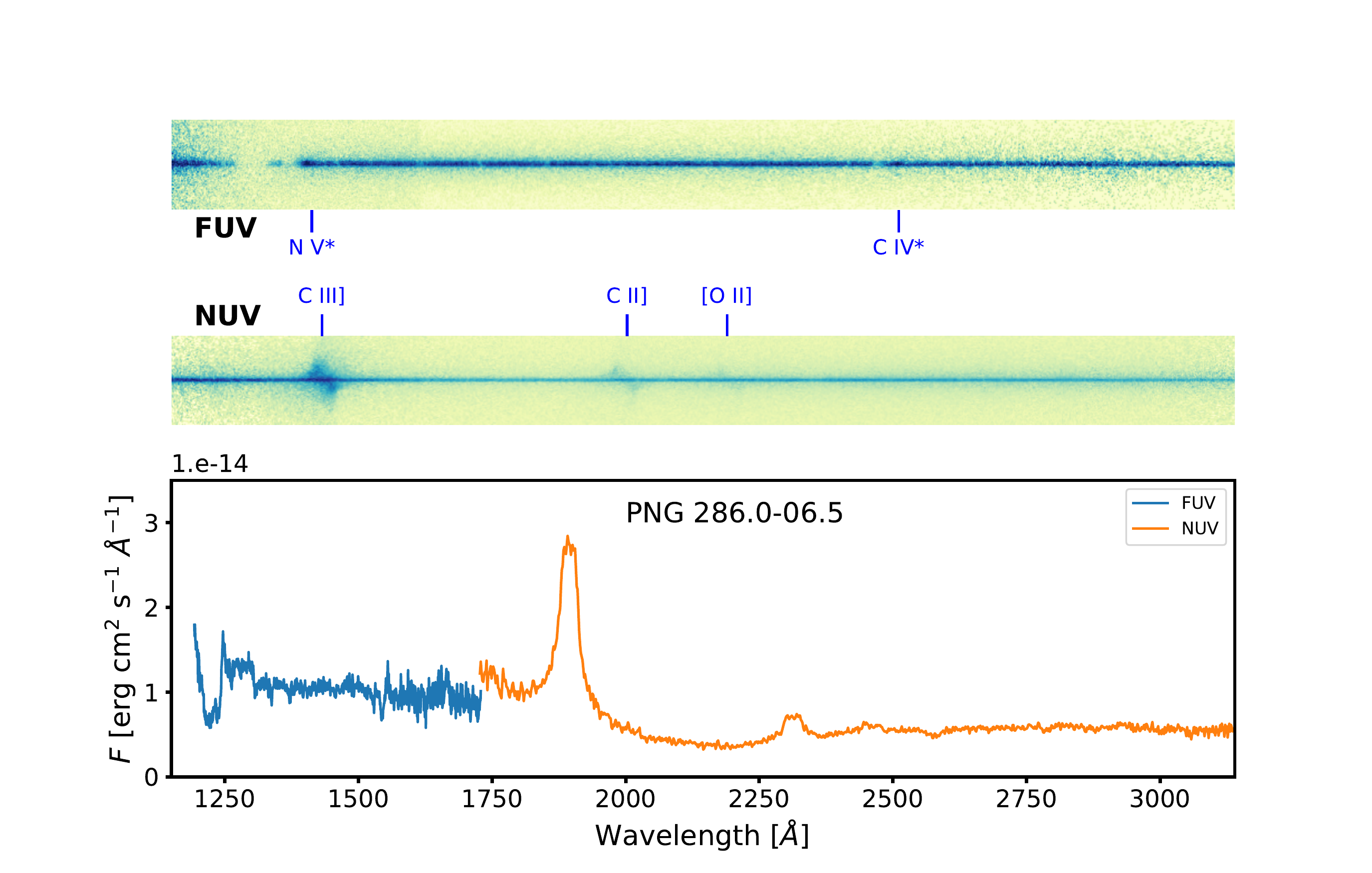}
\caption{False-color rendering of the spectrograms, as in Fig.~1 but for PN~G286.0--06.5 }
\end{figure}

\begin{figure}
   \centering
    \includegraphics[width=0.495\columnwidth]{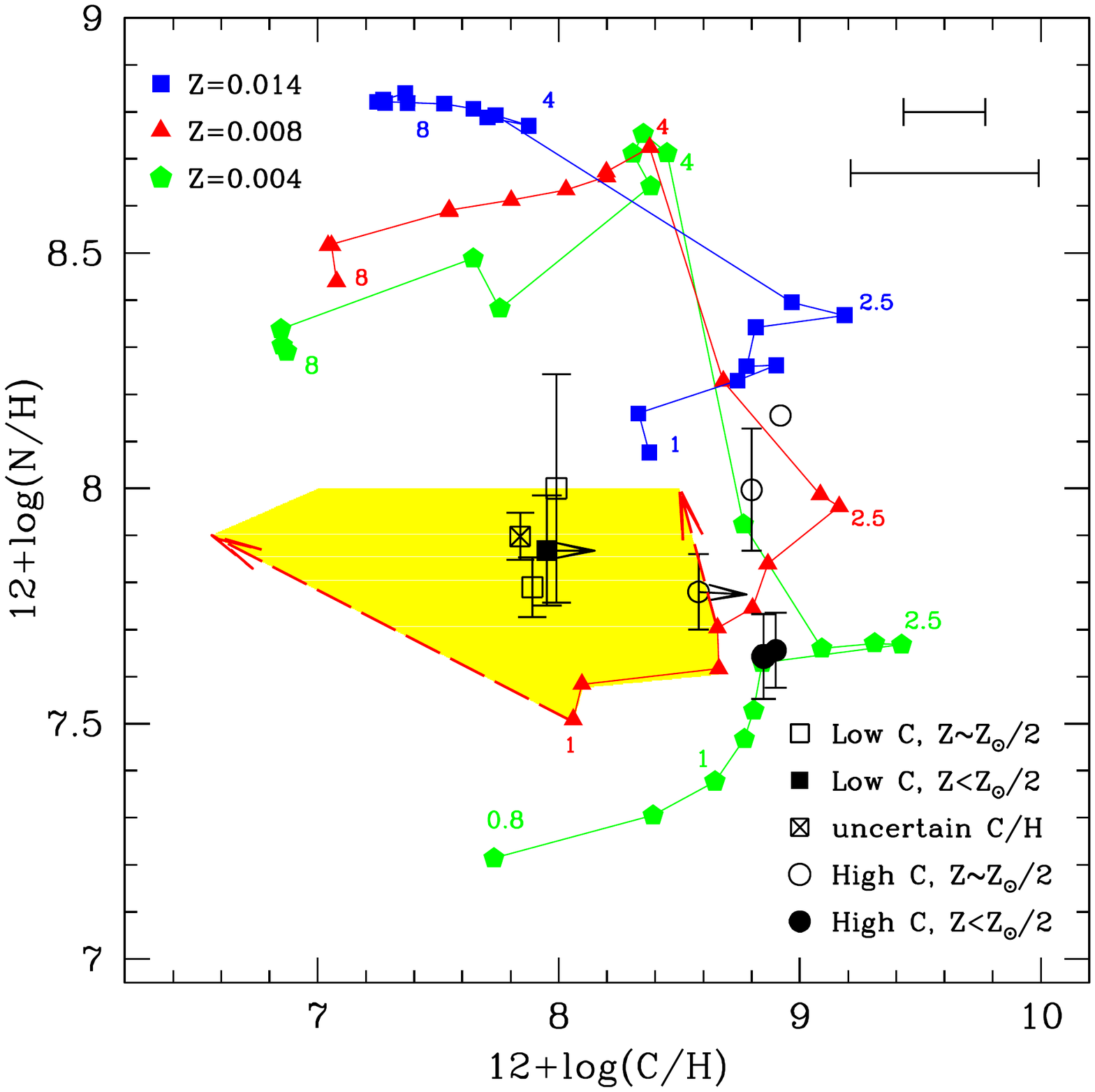}
    \includegraphics[width=0.495\columnwidth]{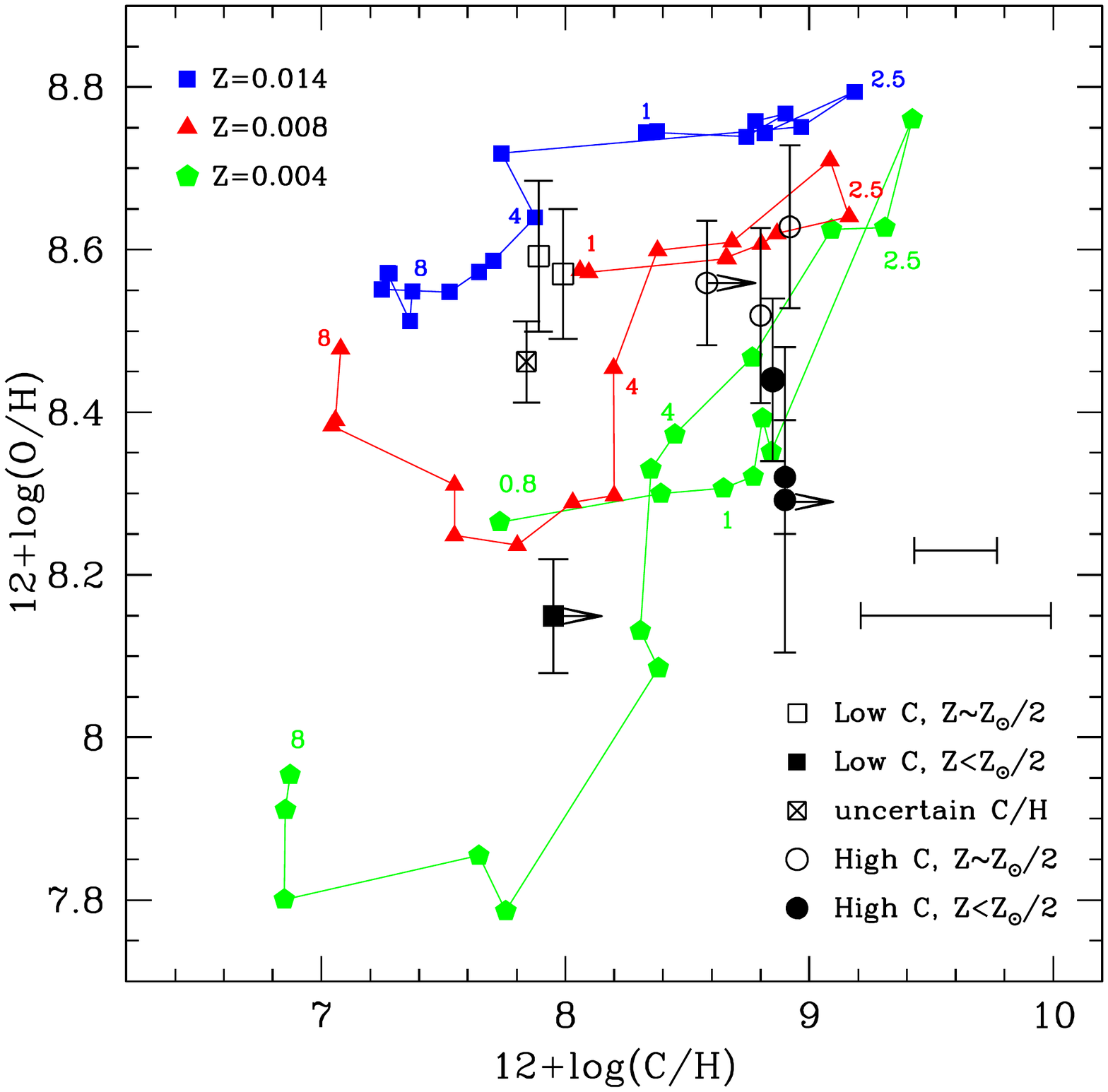}
    \vskip-30pt
    \caption{Left panel: log(N/H)+12 vs. log(C/H)+12; right panel:  log(O/H)+12 vs. log(C/H)+12. In both panels, carbon abundances are from this paper (Table 9, and other abundances are from the literature (Table \ref{tab:Parameters} and references therein).  
 The different black symbols represent the abundance groups from our carbon analysis and interpretation, as given in the legend: Open squares are low-carbon PNe, which we interpret as descending from half-solar metallicity stars; filled squares are low-carbon PNe descending from low metallicity stars; the crossed square represents PN~G053.3+24.0, whose carbon abundance is uncertain.
    The open circles represent enhanced carbon PNe descending from half-solar metallicity stars, whereas filled circles are for enhanced carbon PNe descending from low-metallicity stars (see text).
    We also show in the panel the representative carbon abundance errorbars derived from a 5$\%$ and 10$\%$ uncertainty in the electron temperature, for reference. 
    In both panels, the symbols in color, connected by lines, represent the final abundances of AGB stars with initial metallicities Z=0.014 (blue squares), Z=0.008 (red triangles), and Z=0.004 (green pentagons), calculated from the models. The numbers indicate the initial mass of the model, in $M_{\odot}$. 
    In the left panel, the dashed red arrows point to the final C and N abundances of the 1\,M$_{\odot}$ (left arrow) and 1.5\,M$_{\odot}$ (right arrow) models, if deep mixing in RGB is taken into account. The yellow area in the left panel highlights the range of final C and N values of the ejecta from stars with mass 1\,M$_{\odot}< $M$ < 1.5\,$M$_{\odot}$, if deep mixing in the RGB is taken into account.}
    \label{fig:CNCO}
\end{figure}
\clearpage

\begin{figure}
\includegraphics[width=0.75\columnwidth]{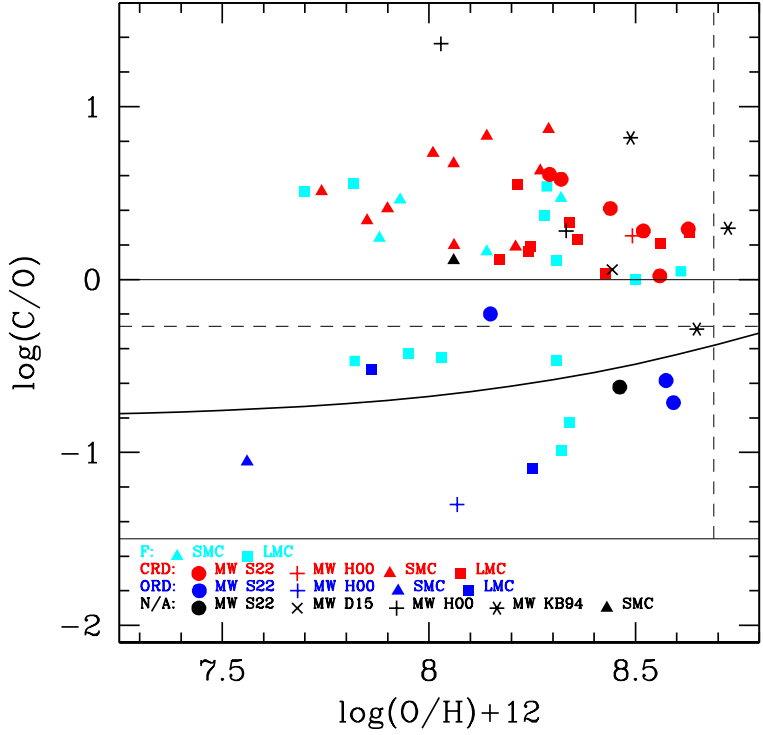}
\caption{Compact PNe in 3 galaxies, in the log(C/O) vs. log(O/H)+12 plane. This plot includes all SMC, LMC, and compact Galactic PNe with $\theta<5\arcsec$, whose carbon abundances has been measured from the UV emission lines.
Symbol shapes represent the PN sample: Triangles: SMC PNe, with abundances from \citet{VSD17} and references therein; squares: LMC PNe, with abundances from \citet{VSD16} and referenced therein; circles: compact Galactic PNe from this work, with carbon abundances from Table 9 and oxygen abundances from table 2; crosses: compact Galactic PNe, with abundances from \citet{2015ApJ...803...23D} (D15); plus signs: compact Galactic PNe, with abundances from \citet{2000ApJ...531..928H} (H00), and asterisks: compact Galactic PNe, with abundances from \citet{KB94} (KB94). 
The symbols are color-coded for their dust type according to the analysis of their Spitzer/IRS spectra \citep{SGG07,SGG12}; cyan: featureless (F) IRS spectra; red: carbon-rich dust (CRD), blue: oxygen-rich dust (ORD), and black: no dust information available through Spitzer/IRS analysis.
The thick solid line represents the fit to the locus of stars \citep{Nicholls}; the light solid line represents C=O; the light dashed lines marks the solar values of C and O \citep{Asplund}.}
\end{figure}

\end{document}